\documentclass[10pt, journal, compsoc]{IEEEtran}

\usepackage[T1,hyphens]{url}
\usepackage{hyperref}

\usepackage{booktabs}
\usepackage{xcolor}
\usepackage{siunitx}
\usepackage{multirow}
\usepackage{xspace}
\usepackage{balance}
\usepackage{graphicx}
\usepackage{tabularx}
\usepackage{enumitem}
\usepackage[caption=false]{subfig}
\usepackage[boldmath]{numprint}

\newcommand{\reduceline}{{\vspace*{-0.15cm}}}
\usepackage[noend]{algpseudocode}
\usepackage[linesnumbered%,ruled
			,vlined]{algorithm2e}
\let\oldnl\nl% Store \nl in \oldnl
\newcommand{\nonl}{\renewcommand{\nl}{\let\nl\oldnl}}% Remove line number for one line

\SetKwIF{If}{ElseIf}{Else}{{{if}}}{}{{else if}}{{{else}}}{}
\SetKwFor{For}{{for}}{}{}

\makeatletter
\newcommand{\removelatexerror}{\let\@latex@error\@gobble}
\makeatother

\setlist[itemize,1]{leftmargin=2mm,itemsep=1mm}
\setlist[enumerate,1]{leftmargin=4mm,itemsep=1mm}

\newcommand{\method}[1]{{\textsf{#1}}}

\newcommand{\return}{\vspace{0.1cm}}

\newcommand{\code}[1]{{{\textbf{#1}}}}
\newcommand{\func}[1]{{\textsf{#1}}}
\newcommand{\var}[1]{{\textit{#1}}}

\newcommand{\pp}{{\hspace{0.15cm}}}
\newcommand{\occ}{{\textsf{matches}}}

\newcommand{\myparagraph}[1]{{\vspace{0.2cm}\noindent\normalsize\bf #1.}}
\newcommand{\bb}[1]{{\textbf{\numprint{#1}}}}

\newcommand{\mytablescale}{0.9}

\usepackage{xcolor}

\newcommand{\collection}[1]{\method{#1}}

\newcommand{\dblp}{{\collection{DBLP}}}
\newcommand{\gnames}{{\collection{Geonames}}}
\newcommand{\dbpedia}{{\collection{DBpedia}}}

\newcommand{\lubm}{{\collection{LUBM}}}
\newcommand{\wat}{{\collection{WatDiv}}}
\newcommand{\freebase}{{\collection{Freebase}}}

\newcommand{\bpt}{{\sf{bits/triple}}}

\newcommand{\sq}{{\sf{sec/query}}}
\newcommand{\nt}{{\sf{ns/triple}}}

\newcommand{\spo}{{\method{SPO}}}
\newcommand{\pos}{{\method{POS}}}
\newcommand{\osp}{{\method{OSP}}}
\newcommand{\sop}{{\method{SOP}}}
\newcommand{\ops}{{\method{OPS}}}
\newcommand{\pso}{{\method{PSO}}}

\newcommand{\select}{{\method{select}}}

\newcommand{\access}{{\method{access}}}

\newcommand{\find}{{\method{find}}}
\newcommand{\scan}{{\method{scan}}}

\newcommand{\three}{{\method{3T}}}
\newcommand{\threecc}{{\method{CC}}}
\newcommand{\two}{{\method{2T}}}
\newcommand{\twop}{{\method{2Tp}}}
\newcommand{\twoo}{{\method{2To}}}

\newcommand{\triplebit}{{\method{TripleBit}}}
\newcommand{\hexastore}{{\method{HexaStore}}}
\newcommand{\hdt}{{\method{HDT-FoQ}}}
\newcommand{\bitmat}{{\method{BitMat}}}
\newcommand{\rdfx}{{\method{RDF-3X}}}
\newcommand{\rdfcsa}{{\method{RDFCSA}}}
\newcommand{\ktriples}{{$k^2$-\method{TRIPLES}}}

\newcommand{\ef}{{\method{EF}}}
\newcommand{\pef}{{\method{PEF}}}

\newcommand{\vbyte}{{\method{VByte}}}
\newcommand{\vbytesimd}{{\method{VByte+SIMD}}}
\newcommand{\comp}{{\method{Compact}}}

\begin{document}

\title{Compressed Indexes for\\Fast Search of Semantic Data}

\author{Raffaele Perego --
Giulio Ermanno Pibiri --
Rossano Venturini
\IEEEcompsocitemizethanks{
\IEEEcompsocthanksitem
Raffaele Perego and Giulio Ermanno Pibiri are affiliated to the
High Performance Computing Lab, ISTI-CNR, Area
della Ricerca di Pisa, Via Giuseppe Moruzzi, 1, 56126 Pisa, Italy.\protect\\
E-mail: $\{$raffaele.perego, giulio.ermanno.pibiri$\}$@isti.cnr.it
\IEEEcompsocthanksitem
Rossano Venturini is affiliated to the Department of Computer Science,
University of Pisa, Viale Largo Bruno Pontecorvo 3, 56127 Pisa, Italy.\protect\\
E-mail: rossano.venturini@unipi.it
\IEEEcompsocthanksitem
This work was partially supported by the BIGDATAGRAPES (EU H2020 RIA, grant agreement N\textsuperscript{\b{o}}780751), the ``Algorithms, Data Structures and Combinatorics for Machine Learning'' (MIUR-PRIN 2017), and the OK-INSAID (MIUR-PON 2018, grant agreement N\textsuperscript{\b{o}}ARS01\_00917) projects.
}% <-this % stops an unwanted space
}

\IEEEtitleabstractindextext{%
\begin{abstract}
The sheer increase in volume of RDF data demands efficient solutions for the
\emph{triple indexing problem}, that is to devise a compressed data structure
to compactly represent RDF triples by guaranteeing,
at the same time, fast pattern matching operations.
This problem lies at the heart of delivering good practical performance
for the resolution of complex SPARQL queries on large RDF datasets.
In this work, we propose a trie-based index layout
to solve the problem and introduce two novel techniques to reduce
its space of representation for improved effectiveness.
The extensive experimental analysis, conducted over a wide range of publicly
available real-world datasets, reveals that our
best space/time trade-off configuration
substantially outperforms existing solutions\\
at the state-of-the-art, by taking 30 -- 60\%
less space \emph{and} %, depending on the triple selection pattern,
speeding up query execution by a factor of 2 -- 81$\times$.
\end{abstract}
}

\maketitle

\IEEEdisplaynontitleabstractindextext

\IEEEpeerreviewmaketitle

\IEEEraisesectionheading{\section{Introduction}\label{sec:intro}}
\IEEEPARstart{T}{he} Resource Description Framework (RDF\footnote{\url{https://www.w3.org/RDF}}) is a W3C standard offering a general graph-based model for describing information as a set of (\var{subject}, \var{predicate}, \var{object}) relations, known as triples. Representing data in RDF allows subject and object entities to be unambiguously identified and connected through directed and explainable relationships, thus favoring the integration and reuse of different information sources.
Although RDF was initially conceived as a metadata model for the Semantic Web and the Linked Data~\cite{semWeb}, its generality and flexibility favoured its diffusion in other domains ranging from digital libraries to bioinformatics and business intelligence. Moreover, the success of initiatives such as \emph{schema.org}
and \emph{opengraph}\footnote{\url{http://ogp.me}} made RDF the de-facto standard format for publishing semi-structured information in social networks and Web sites.
In fact, major search engines like Google and Bing are providing increasingly-better support for RDF.

Such wide popularity encouraged the development of several data management systems able to deal with large RDF datasets and the complexity of querying them via SPARQL\footnote{\url{https://www.w3.org/TR/sparql11-overview}},
a query language that understands the RDF model and allows to select and join graph-structured data based on both content and patterns.
Not surprisingly, the increasing volume of RDF data available on-line and in various repositories pushed researchers to investigate specific solutions enabling users and software agents to store, access and query RDF graph-structured data efficiently.
In this direction we can identify four relevant research topics.

\begin{itemize}
\item
%$\cdot$
\emph{Compressed string dictionaries.} Each RDF statement has three components: a \var{subject} (S), an \var{object} (O), and a \var{predicate} (P, also called a \var{property}) that denotes a relationship. Each one of these components is a URI string (or even a literal in the case of an object). Since URI strings can be very long and the same URI generally appears in many RDF statements, the components of triples are commonly mapped to integer IDs to save space, so that each triple in the dataset can be represented with three integers. 

\item
%$\cdot$
\emph{Triple indexing data structures.} Indexes built over the set of triples should allow fast access to data for processing complex SPARQL queries involving large sequences of \emph{triple selection patterns} over RDF graphs \cite{RDFSurvey,RDFSurvey2}.

\item
%$\cdot$
\emph{Query-planning algorithms.} An effective query-planning algorithm has to find a suitable order to the set of atomic selection patterns that are needed to solve a SPARQL query, in order to speed up its execution and optimize expensive join operations~\cite{RDF3x,RDF3xjournal,TripleBit}.

\item
%\noindent$\bullet$
\emph{Inference.} RDF triples are used to infer new relationships
in order to improve the quality of the data and discover possible
inconsistencies~\cite{paulheim2013type,subercaze2016inferray}.
\end{itemize}

\noindent
In this work, we focus on the \emph{triple indexing problem}
that is to design a static index for the integer triples
that attains to efficient resolution of all possible selection patterns
using as little space as possible.
This is crucial to guarantee practical SPARQL query evaluation.
Therefore, we do not directly manage
a string dictionary mapping URIs to integer IDs that,
as discussed above, is a different problem.

Moving from a critical analysis of the state-of-the-art, we note that
existing solutions to the problem
require too much space, because either:
rely on materializing all possible
permutations of the S-P-O components~\cite{Hexastore,RDF3xjournal};
use expensive additional supporting structures~\cite{HDT2,RDF3xjournal};
do not use sophisticated data compression techniques to effectively reduce
the space for encoding triple identifiers~\cite{BitMat,TripleBit}.
Furthermore, this additional space overhead does not always pay off in terms of
reduced query response time.
The aim of this work is that of addressing these issues
by proposing compressed indexes for RDF data that are both compact \emph{and} fast.

\myparagraph{Our contributions} In particular, our detailed contributions are as follows.
\begin{enumerate}

\item We propose the use of a \emph{trie}-based index layout delivering a considerably better efficiency for all triple selection patterns, thanks to the cache-friendly nature of its pattern matching algorithm.
Specifically, the index materializes three different permutations of the triples in order to (symmetrically) support all
triple selection patterns with one or two wildcard symbols.
By leveraging on well-engineered compression techniques,
% to compactly represent the indexed triples and maintain efficient query time,
we show that this design is already as compact as
the most space-efficient competitor in the literature and 2 -- 4$\times$ faster on average for all selection patterns.

\item Starting from the aforementioned index layout, we devise two optimization aimed at reducing the redundancy of the representation. The first technique builds on the observation that the order of the triples given by a permutation can be actually
exploited to compress another permutation, hence \emph{cross-compressing} the index.
The second technique shows that it is possible to \emph{eliminate} a permutation without
affecting (or even improving) triple retrieval efficiency.

\item Extensive and thorough experiments aimed to assess the space and time performance of our proposal versus state-of-the-art competitors are conducted on publicly available RDF datasets with a number of triples ranging from 88 millions up to 2 billions and show that our best space/time trade-off configuration
substantially outperforms
existing solutions at the state-of-the-art, by taking 30 -- 60\%
less space \emph{and}
speeding up query execution by a factor of 2 -- 81$\times$.
\end{enumerate}

\myparagraph{Source code}
In the interest of reproducibility,
our code is
available at \url{https://github.com/jermp/rdf_indexes}.

\section{Related work}\label{sec:related}

In the last decade many researchers investigated RDF management systems from different perspectives and a complete review of the efforts in this field is out of the scope of this paper. Readers interested in the general topic of RDF data management can refer to two very recent surveys~\cite{RDFSurvey,RDFSurvey2}.
In the following we focus on the works proposing indexing structures built over the set of triples to support efficient RDF query processing.
These works usually exploit string dictionaries to compactly code the URIs occurring in the triples with unique integer IDs and adopt specific domain-dependent techniques to index various S-P-O permutations in order to enhance locality in the access to all the triples matching a given selection pattern: {\spo}, {\osp}, {\sop}, {\ops}, {\pos} and {\pso}.
Depending on the solution proposed, some or all these permutations are materialized and sorted by the values of the IDs in the three columns.
% Then a distinct index is built for each of them.

A simple incarnation of this approach is called
\emph{vertical-partitioning}~\cite{AbadiMMH09},
where the permutation {\pso} is materialized.
In particular, a 2-column table is built for each predicate and it lists
all the (\var{subject}, \var{object}) pairs sorted on the \var{subject}
component to permit fast search and good compression effectiveness.
Vertical-partitioning can be generalized to any permutation of
the S-P-O components. For example, instead of materializing {\pso},
the triples can be partitioned by the \var{subject} component
and the (\var{predicate}, \var{object}) pairs stored in sorted tables.
In the extreme case, all the six possible
permutations can be materialized.

{\hexastore}~\cite{Hexastore} and {\rdfx}~\cite{RDF3x,RDF3xjournal}
follow this exhaustive indexing strategy.
They both build and materialize six indexes, one for each possible permutation of the three RDF components.
{\rdfx} introduces also additional indexes over count-aggregated variants for all three two-dimensional and all three one-dimensional projections. The great flexibility of these solutions at querying time is paid with a very large space occupancy. To partially address this issue, inspired by compression methods for inverted lists in text retrieval systems, {\rdfx} exploits {\vbyte} to encode the delta gaps computed from the increasingly ordered sequences of integers stored in the clustered B+ trees used for the indexes.
The advantage of such organization is that any selection pattern with wildcards possibly occurring in one or two components among the subject, predicate, or object can be efficiently processed on the most suitable index, e.g., the one where the matching triples are stored contiguously in memory. Moreover, also joins for processing structural SPARQL queries are efficiently supported by fast triple selection over two indexes.
However, the clear disadvantage is the additional space occupancy and overhead to store and manage the redundant information.

A completely different approach relies on representing the triples in memory with bitmaps~\cite{BitMat,TripleBit}.
{\bitmat}~\cite{BitMat} encodes the RDF data with a 3D bit-cube, where the three dimensions correspond to S, P and O, respectively.
%A bit set to 1 in a cell denotes the existence of a unique triple formed by the combination of the RDF components identified by the values of the three coordinates. By slicing the 3D bit-cube along a dimension several two dimensional bit-matrices are obtained.
%By slicing for example along the predicate dimension SO {\bitmat}s are obtained  that are then inverted to originate the OS ones. Analogously the bit-cube is sliced along the subject and object dimensions to build PO and PS {\bitmat}s, respectively. 
%The rows of these bit-matrices are then compressed by explicitly coding the lengths of the runs of consecutive 0s or 1s.
{\triplebit}~\cite{TripleBit} encodes triples with a bit-matrix. In this matrix each column represents a distinct triple where only two bits are set to 1 in correspondence of the rows associated with the subject and the object of the triple. Symmetrically, each row corresponds to a distinct entity value occurring as subject or object in the triples uniquely identified by the bits set to 1 in the row.
Since the resulting bit matrix is very sparse it is compressed at the column level by simply coding  the position of the  two rows corresponding to the bits set, i.e., the identifiers of the subject and object of the associated triple.
%In addition, columns are sorted by predicate to build for each predicate a disjoint vertical partition of the bit matrix. Partitions are then organized in chunks of fixed size and auxiliary indexes are provided to access efficiently all the partitions and chunks matching a given subject or object.
The experimental assessment discussed in the paper shows that {\triplebit} outperforms {\rdfx} and {\bitmat} by up to 2 -- 4 orders of magnitude on large RDF datasets.
Some work also investigated the impact of
graph processing units (GPUs) to process binary matrices~\cite{sankar2014efficient}.
Nevertheless, the space occupancy and scalability of such techniques is not very good.

Other authors have explored how variations of well-known data structures like $k^2$-trees~\cite{brisaboa2009k} and the Sadakane's \emph{compressed suffix array} (CSA)~\cite{sadakane2003new} can be exploited to compactly represent RDF datasets.
In particular, {\ktriples}~\cite{alvarez2015compressed} partitions the dataset into disjoint subsets of (\var{subject}, \var{object}) pairs, one per predicate, and represents the (highly) sparse bit matrices with $k^2$-trees.
Another approach called {\rdfcsa}~\cite{brisaboa2015compact,cerdeira2016self} builds an integer CSA index~\cite{farina2012word} over the sequence of concatenated triple IDs, with the use of truncated Huffman codes on integer gaps and run lengths for optimized performance.
The experimental assessment shows that {\rdfcsa} requires roughly twice the space of the {\ktriples} but it is up to two order of magnitude faster than the former.
Although both these solutions outperform existing solutions like {\rdfx} in both space usage and query efficiency, no implementation is publicly available to reproduce their results\footnote{Personal communication.}.
%However, the fastest solution discussed in the papers requires several microseconds per matched triple for all selection patterns, whereas in this paper we describe solutions taking fractions of a microsecond per triple.

The RDF index most similar to our solution is {\hdt} (\emph{Focused on Querying})~\cite{HDT1,HDT2}, a trie-based solution that exploits the skewed structure of RDF graphs to reduce space occupancy while supporting fast querying. The {\hdt} format includes
%\footnote{\url{https://www.w3.org/Submission/2011/03/}}
% {\hdt} has been proposed to W3C as a compact binary serialization format and 
 a \emph{header}, detailing logical and physical metadata, the \emph{dictionary}, encoding all the unique entities occurring in the triples as integers, and the set of \emph{triples}  encoded in a single {\spo} trie data structure.
 %The triples are grouped by subject, ordered by subject ID and represented compactly by considering a different tree for each distinct subject. 
 %Due to the ordering employed the subject of each group of triples is implicitly defined and its representation is omitted in the adjacency list used for coding the tree.
%\giulio{Also see this: \url{http://www.rdfhdt.org/hdt-binary-format/}}
% The \spo\ trie is used to solve 5 selection patterns out of 8. %Also S?O is solved on {\spo}, just because they can not do otherwise (not because they recognize is faster, so it is just a matter of luck).
%The pointers in the trie are in unary codes with additional rank/select succinct data structures for indexing.
In order to support predicate-based retrieval, the second level of the trie is represented with a \emph{wavelet tree}~\cite{grossi2003}.
%This allows ?PO to be solved by determining all occurrences of a given predicate p and searching for the given o among the children of p.  The ?P? selection pattern is supported similarly.
Finally, additional inverted lists are maintained for object-based triple retrieval. In particular, for each object $o$, an inverted list is built, listing all pairs (\var{subject}, \var{predicate}) of the triples that contain $o$. Thus, searches are carried out by accessing each pair and searching for it in the trie.
A similar approach based on wavelet trees was also
proposed by Cur{\'{e} et al.~\cite{CureBRF14}.

As we are going to detail next, our base solution is similarly based on the trie data structure but instead of maintaining a single trie we exploit efficient and effective compression strategies resulting in a faster solution for all the triple selection patterns and a smaller space occupancy. 

%\myparagraph{Differences between {\hdt} and {\two}}
%\giulio{this paragraph to be removed later, using informal words in the following.}
%We initially present the layout with three trie (compact, EF-single, PEF, VByte) and show that this solution already matches the space of {\hdt} but it is much faster.
%Then we refine our solution with the two-trie approach. In particular, we maintain the permutation {\spo} to solves 5 selection patterns out of 8, by motivating that S?O can be solved on {\spo}. Since we do not use a wavelet-tree on P, it is faster than the algorithm used by {\hdt}.
%We explain the efficiency of the algorithm in detail.
%In order to support ?PO and ??O, we maintain another permutation OPS that is the reverse of SPO. This is much better because it solves these two selection patterns optimally by find+scan, thus eliminating the (many) rank/select operations needed by {\hdt}.
%Finally, to support ?P?, we can either proceed by scanning the predicates or keeping the mapping $p$ $\rightarrow$ \emph{list of subjects in the triples having predicate p}, noting that the approach is competitive and we match the performance of {\hdt} because predicates are few in number. Put performance counts for this algorithm too.
%\return

\section{The permuted trie index}\label{sec:permuted}
In this section we introduce our trie-based index that solves the \emph{triple indexing problem} mentioned in Section~\ref{sec:intro}: compressing a large set of integer triples by granting the efficient resolution of sequences of triple selection patterns. In particular, in Section~\ref{subsec:core} we introduce the base indexing data structure and in Section~\ref{subsec:remapping} and~\ref{subsec:twotries} we discuss two variants aimed at reducing redundancies in the representation.

In order to better support our design choices and explain the intuition behind the described ideas, we show in the following the results of some motivating experiments conducted on the {\dbpedia} dataset -- ``the nucleus for a Web of Data''~\cite{auer2007dbpedia} -- that is the English version of DBpedia (version 3.9) containing more than 351 millions of triples.
(See also Table~\ref{tab:datasets} at page~\pageref{tab:datasets}).
In Section~\ref{sec:experiments} we will report on the comprehensive set of experiments conducted to assess the  performance in space and time of our implementations versus state-of-the-art competitors on publicly available RDF datasets of varying size and characteristics.
%Refer to Table~\ref{tab:datasets} for the detailed statistics about the {\dbpedia} and the other datasets used.

\subsection{Data structure}\label{subsec:core}
As a high-level overview, our index maintains three different permutations of the triples, with each permutation sorted to allow efficient searches and effective compression.
The permutations chosen are {\spo}, {\pos} and {\osp} in order to (symmetrically) support all the six different triple selection patterns with one or two wildcard symbols: \textsf{SP?} and \textsf{S??} over {\spo}; \textsf{?PO} and \textsf{?P?} over {\pos}; \textsf{S?O} and \textsf{??O} over {\osp}. The two additional patterns with, respectively, all symbols specified or none, can be resolved over any permutation, e.g., over the canonical {\spo} in order to avoid permuting back each returned triple.

\begin{figure}[t]
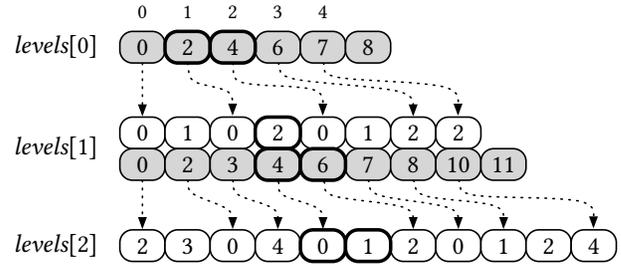

\centering
\includegraphics[scale=0.6]{{{imgs/trie}}}
\caption{A trie data structure representing a set of triples. Shaded boxes indicate pointers whereas the others refer to the nodes of the trie.
Nodes in the first level are implicit, thus are not part of the data structure but reported
here in smaller font for better visualization.
Similarly, the dashed arrows are just for representational purposes and point to the position written in the corresponding originating box.
Lastly, we highlight in thick stroke the nodes and pointers that are accessed during the resolution of the pattern $(1,2,\textsf{?})$.
}
\label{fig:trie}
\end{figure}

Each permutation of the triples is represented as a 3-level trie data structure, with nodes at the same level concatenated together to form an integer sequence. We keep track of where groups of siblings begin and end in the concatenated sequence of nodes by storing such pointers as absolute positions in the sequence of nodes. Therefore, the pointers are integer sequences as well.
Moreover, since the triples are represented by the trie data structure in sorted order, the $n$ node IDs in the first level of each trie are always complete sequences of integers ranging from 0 to $n-1$ and, thus, can be omitted.
We can model each trie data structure with an array $\var{levels}[0,1,2]$ of three objects, each one having two integer sequences of \var{nodes} and \var{pointers}. An exception is represented by $\var{levels}[0]$ for which, as discussed above, nodes are missing, and by $\var{levels}[2]$ for which pointers are missing.

Refer to Fig.~\ref{fig:trie} for a pictorial example in which the following set of
triples is indexed: $\{(0,0,2)$, $(0,0,3)$, $(0,1,0)$, $(1,0,4)$, $(1,2,0)$, $(1,2,1)$, $(2,0,2)$, $(2,1,0)$, $(3,2,1)$, $(3,2,2)$, $(4,2,4)\}$.

The advantage of the introduced layout is \emph{two-fold}.
First, we can effectively compress the integer sequences that constitute the levels of the tries to achieve small storage requirements.
Second, as exemplified above, the triple selection patterns are made
\emph{cache-friendly} and, hence, efficient by requiring to simply \emph{scan} ranges of consecutive nodes in the trie levels.
In what follows, we explore and quantify the impact of these two advantages.

Before continuing,
an important consideration is in order.
The described data structure is \emph{static}, i.e.,
it does \emph{not} directly support dynamic updates.
Note, however, that a simple amortized solution could
solve this limitation.
For example, we could maintain a ``small'' index
holding the most recent updates.
Whenever the small index reaches a predefined size,
its content is merged with the one of the
main, static, index.
Queries also need to involve both indexes and their results have to be merged 
accordingly.

\myparagraph{Solving triple selection patterns}
The pseudo code reported in Fig.~\ref{alg:select} illustrates how triple patterns with one or two wildcard symbols are supported by our index.
Given a sequence \var{S}, function
$\var{S}.\func{find}(\var{i},\var{j},\var{x})$
finds the ID \var{x} in the range
$\var{S}[\var{i},\var{j})$ and returns its absolute position in the sequence. If \var{x} is not found in the range, a default position, e.g., $-1$,
is returned to signal the event and the number of matches will be 0.
% (for simplicity, in the pseudo code, we assume the \var{triple} to be valid, i.e., indexed by the data structure).
Function $\var{S}.\func{iterator\_at}(\var{i})$ instead instantiates an iterator
starting at $\var{S}[\var{i}]$.
We assume that \func{invalid\_iterator} is a function returning
an iterator over an \emph{empty} range (that is invalid).
%We assume that accessing the pointers at position \var{i} fetches a pair of adjacent values (\var{begin}, \var{end}), such that \var{begin} = \var{pointers}[\var{i}] and \var{end} = \var{pointers}[$\var{i}+1$] with the value $\var{end}-\var{begin}$ giving the number of children of the node at position \var{i} in the previous level of the trie (line 3 and 9).
Furthermore, the \func{select} algorithm creates two iterators to scan ranges of the second and third levels of the trie, respectively. These iterators are then used to define a final iterator that combines the iterating capabilities of both objects (line 20 of the pseudo code).

For example, pattern matching $(1,2,\textsf{?})$ will return the two triples $(1,2,0)$ and $(1,2,1)$ because 
these are the ones sharing the first two specified components $(1,2)$.
Fig.~\ref{fig:trie}
highlights the nodes and pointers accessed during the resolution of such pattern.
In this case, we begin by fetching the pair of pointers $(2,4) = (\var{levels}[0].\var{pointers}[1], \var{levels}[0].\var{pointers}[2])$ (lines 5 and 6).
Next, we have to find the position of the ID 2 among the nodes in the second level. We do this with 3 = $\var{levels}[1].\var{nodes}.\func{find}(2,4,2)$ (line 9).
Given that position, we fetch a new pair of pointers $(4,6) = (\var{levels}[1].\var{pointers}[3], \var{levels}[1].\var{pointers}[4])$
(lines 16 and 17).
Finally, we know that all completions of the prefix $(1,2)$
are given by the node IDs
found in the range $\var{levels}[2].\var{nodes}[4,6)$, that are $0$ and $1$.
These will be returned by the \func{iterator} object created in line 20 of the pseudo code.

We now discuss the time complexity of a triple selection pattern.
We use the following nomenclature: $n$ indicates the total number
of triples; {\occ} indicates the number of
matches for a given pattern;
$|\textsf{S}|$, $|\textsf{P}|$ and $|\textsf{O}|$ indicate
the number of distinct subjects, predicates and objects
respectively.
Given an integer sequence \var{S},
we assume that: (1) we can randomly \func{access} any position of \var{S}
and retrieve the integer at such position in $O(1)$ time;
(2) the complexity of
instantiating an iterator over the range $\var{S}[i,j)$
and returning every integer in such range is $\Theta(j - i + 1)$,
that is linear in the size of the range.
Therefore it follows that the
$\var{S}.\func{find}(\var{i},\var{j},\var{x})$
operation can be implemented using
binary search in $O(1 + \log(j - i))$
that is $O(1 + \log|S|)$ time for any $x$ and interval.
It is then straight forward to see that the pattern
\textsf{???} is supported in $\Theta(n)$ time,
that is $\Theta(1)$ per triple.
The patterns with two wildcard symbols are supported
in $\Theta(1 + \occ)$ time.
%that is $\Theta(1 + |\textsf{S}|)$,
%$\Theta(1 + |\textsf{P}|)$ and $\Theta(1 + |\textsf{O}|)$ time
%for \textsf{S??}, \textsf{?P?} and \textsf{??O} respectively.
The patterns with one wildcard need one
\func{find} operation to be resolved in the second
level of the trie dedicated to their support.
Therefore, \textsf{SP?} takes $O(1 + \log|\textsf{P}| + \occ)$ time,
\textsf{S?O} and \textsf{?PO} take $O(1 + \log|\textsf{S}| + \occ)$
and $O(1 + \log|\textsf{O}| + \occ)$ time respectively.
Finally, the pattern \textsf{SPO} needs two \func{find}
operations, thus taking $O(1 + \log|\textsf{P}| + \log|\textsf{O}|)$ time.
\return

%Apart from the cost of the \func{find} operation called for patterns with one wildcard symbol and the time needed to access the pointer sequences, the time spent per matched triple is $O(1)$, that is optimal.

%Next, we illustrate how these query patterns are efficiently
%supported over the compressed node sequences.

%how the \func{find} operation and the random \func{access} can be efficiently supported directly over the compressed node sequences.

\begin{figure}[t]
\removelatexerror
\scalebox{\mytablescale}{
    \input{select.tex}
}

\caption{Function \func{select}  solving triple selection patterns with one or two wildcard symbols. The input is a \var{triple} object, assumed to be formed by its \var{first}, \var{second} and \var{third} attributes.
\label{alg:select}}
\end{figure}

\begin{table*}[t]
\centering
\caption{Performance of various compressors applied to the sequences of \emph{nodes} of all tries and levels for the {\dbpedia} dataset. Performance is reported in {\bpt} and in nanoseconds per integer when performing {\access}, {\find} and {\scan} operations. In parentheses we report the percentage of space occupancy taken out by the specific sequence with respect to the whole index.}
    \scalebox{\mytablescale}{
        \hspace{-0.35cm}
%\npdecimalsign{.}
%\nprounddigits{2}
\begin{tabular}{
c
l
c
%    S[table-format=.2,round-precision=2]
%    S[table-format=.2,round-precision=1]
%    S[table-format=.2,round-precision=0]
%    S[table-format=.2,round-precision=0]
r@{\hspace{2pt}}
r
r
r
c
c
%    S[table-format=.2,round-precision=2]
%    S[table-format=.2,round-precision=1]
%    S[table-format=.2,round-precision=0]
%    S[table-format=.2,round-precision=0]
r@{\hspace{2pt}}
r
r
r
c
c
%    S[table-format=.2,round-precision=2]
%    S[table-format=.2,round-precision=1]
%    S[table-format=.2,round-precision=0]
%    S[table-format=.2,round-precision=0]
r@{\hspace{2pt}}
r
r
r
c
}

\toprule

&
&
& \multicolumn{5}{c}{\spo}
&
& \multicolumn{5}{c}{\pos}
&
& \multicolumn{5}{c}{\osp}
\\

\cmidrule(lr){4-8}
\cmidrule(lr){10-14}
\cmidrule(lr){16-20}

&
&
& \multicolumn{2}{c}{\bpt}
& {\access}
& {\find}
& {\scan}
&
& \multicolumn{2}{c}{\bpt}
& {\access}
& {\find}
& {\scan}
&
& \multicolumn{2}{c}{\bpt}
& {\access}
& {\find}
& {\scan}
\\

\midrule
\multirow{4}{*}{\rotatebox[origin=c]{90}{Level 2}}
& {\comp}
&
& 4.74
& \small{(4.41\%)}
& 1.6
& 22
& 3
&
& 10.42
& \small{(9.69\%)}
& 1.4
& 285
& 2
&
& 22.15
& \small{(20.61\%)}
& 2.2
& 63
& 4
\\

& {\ef}
&
& 2.15
& \small{(2.33\%)}
& 24.4
& 74
& 4
&
& 4.54
& \small{(4.93\%)}
& 18.6
& 533
& 3
&
& 21.37
& \small{(23.19\%)}
& 34.0
& 168
& 5
\\

& {\pef}
&
& 2.01
& \small{(2.51\%)}
& 65.9
& 83
& 5
&
& 1.34
& \small{(1.67\%)}
& 73.1
& 200
& 4
&
& 18.06
& \small{(22.58\%)}
& 79.8
& 135
& 6
\\

%& {\vbyte}
%&
%& 3.79
%& \small{({\e}\%)}
%& 108.9
%& 115
%& 3
%&
%& 3.42
%& \small{({\e}\%)}
%& 261.4
%& 13008
%& 2
%&
%& 20.40
%& \small{({\e}\%)}
%& 606.9
%& 1084
%& 6
%\\

& {\vbytesimd}
&
& 3.79
& \small{(4.07\%)}
& 68.2
& 77
& 2
&
& 3.42
& \small{(3.67\%)}
& 154.3
& 13067
& 1
&
& 20.40
& \small{(21.90\%)}
& 429.3
& 905
& 5
\\

\midrule
\multirow{4}{*}{\rotatebox[origin=c]{90}{Level 3}}
& {\comp}
&
& 27.00
& \small{(25.12\%)}
& 1.6
& 31
& 4
&
& 25.00
& \small{(23.26\%)}
& 2.6
& 69
& 4
&
& 11.00
& \small{(10.24\%)}
& 2.2
& 8
& 5
\\

& {\ef}
&
& 26.90
& \small{(29.19\%)}
& 31.1
& 97
& 6
&
& 24.17
& \small{(26.23\%)}
& 41.6
& 185
& 5
&
& 5.86
& \small{(6.36\%)}
& 30.4
& 60
& 6
\\

& {\pef}
&
& 26.36
& \small{(32.95\%)}
& 77.0
& 115
& 6
&
& 20.31
& \small{(25.39\%)}
& 90.0
& 150
& 6
&
& 4.76
& \small{(5.95\%)}
& 67.5
& 97
& 7
\\

%& {\vbyte}
%&
%& 26.57
%& 554.8
%& 553
%& 6
%&
%& 22.95
%& 606.9
%& 1085
%& 6
%&
%& 8.87
%& 121.1
%& 121
%& 5
%\\

& {\vbytesimd}
&
& 26.57
& \small{(28.52\%)}
& 418.8
& 421
& 6
&
& 22.95
& \small{(24.64\%)}
& 431.6
& 910
& 5
&
& 8.87
& \small{(9.52\%)}
& 71.8
& 75
& 5
\\

\bottomrule

\end{tabular}
    }
\label{tab:compressors}
\end{table*}

\myparagraph{Supporting range queries}
Another relevant characteristic of the introduced layout is
that it can support also \emph{range queries}, i.e., queries
filtering the set of triples to be returned by means of
range constraints.
For example, we could impose a limit on the objects of the
pattern \textsf{?PO} by requesting all subjects with a given
property \emph{and} having their objects $o$ such that
$\ell < o < r$, with $\ell$ and $r$ being two fixed values.

%All that is required is a simple modification of the assignment
%of URIs to integer identifiers (ids).
%Instead of using the standard lexicographic assignment, i.e.,
%all strings are sorted and consecutive ids are assigned in such
%order,

In order to support such queries we can modify the default lexicographic assignment of URIs
to IDs as follows.
Strings still follows a lexicographic ID assignment, i.e., these are sorted
lexicographically and consecutive IDs are assigned in such order.
Numeric types, instead -- the ones over which we could possibly
express a range restriction --  such as integers or 
real numbers, dates,
and so on\footnote{\url{https://www.w3.org/TR/rdf-sparql-query/\#operandDataTypes}.},
are sorted in increasing order and compressed in a distinct data structure, say $R$.
This $R$ data structure is just a sorted integer sequence that,
as we are going to illustrate next,
supports binary search directly over the compressed representation.

Now, the range restriction $\ell < \texttt{?value} < r$ will
be handled as follows.
\begin{enumerate}
\item \emph{The lower and upper bounds, $\ell$ and $r$, are searched in $R$
to obtain their IDs, say $id_{\ell}$ and $id_{r}$. In the case  $\ell$  ($r$) is not present in $R$, let $id_{\ell}$ ($id_{r}$) be the ID of the closest value in $R$ smaller than $\ell$ (larger than $r$).}
\item \emph{All entities whose IDs are larger than $id_{\ell}$ and less than $id_{r}$
will bound to the variable \textup{\texttt{?value}}
and returned using the  algorithm  described in Fig. 2.}
\end{enumerate}
In conclusion, range queries need \emph{at most two} additional searches
into a separate
data structure with respect to a \textsf{select} query.
As we will see in  Section~\ref{sec:experiments},
the space cost of $R$ is small because its data is sorted
and very compressible.
However, we do not focus much on range queries in the rest of the paper, hence
we assume a traditional lexicographic ID assignment in the following.

\myparagraph{Representation}
A key characteristic of the trie data structure is to conceptually replace runs of the same ID $x$ in a sequence with the pair $(\var{x}, \var{pointer})$,
where the \var{pointer} information indicates the run length and where the run is located in the sequence. This already produces significant space savings when triples share many repeated IDs, as it holds for large RDF datasets.
Again, refer to Fig.~\ref{fig:trie} for a basic graphical evidence of this fact.
We note that the sequences of \var{nodes} and \var{pointers} on each trie level can be effectively compressed using a wealth of available techniques, developed to compactly represent (long) integer sequences with excellent search capabilities, e.g., the ones appearing in inverted indexes~\cite{survey}.
A review and discussion of all such techniques is out of scope for the contents of this paper: we briefly overview the ones relevant for our purposes
and
point the interested reader to the more general survey by Zobel and Moffat~\cite{zobel2006inverted} and the ones by Pibiri and Venturini~\cite{pibiri2018inverted,survey}
for a complete and more detailed description.
%discussing the most recent developments.

Therefore, the problem we face now is the one of compactly representing the levels of each trie by achieving, at the same time, an efficient execution of \func{select} algorithm that, besides the \func{find} function, demands fast random {\access} to the pointer sequences (see Fig.~\ref{alg:select}).
Below we discuss some different options to achieve this goal, chosen as representative of the many available.

A first (obvious) option would be to minimize the number of bits needed to encode an integer, taking $\lceil \log_2 \var{max} \rceil$ bits per value with \var{max} being the maximum one in the sequence.
The advantage of this technique -- indicated as {\comp} in the following -- lies in its simplicity and efficiency, given that a few bit-wise operations are needed to implement the random {\access} operation and, consequently, the $\func{find}(\var{i}, \var{j}, \var{x})$ operation that can be supported by binary searching
the range $\var{S}[\var{i}, \var{j})$.

Instead, an elegant integer encoder fulfilling the {\select} requirements is \emph{Elias-Fano} (\ef)~\cite{elias,fano}.
Elias-Fano combines a fast, namely constant-time, random {\access} algorithm with space close to the information theoretic minimum.
In particular, the number of bits needed to represent a sorted integer sequence
$\var{S}[0,n)$ is at most $n\lceil\log(m/n)\rceil + 2n$, where
$m \geq S[n-1]$ is the universe of representation of the sequence.
The \emph{partitioned} variant ({\pef}), introduced by
Ottaviano and Venturini~\cite{ottaviano2014partitioned},
reduces the space of Elias-Fano by breaking a sequence into
partitions and encoding each partition separately.
%Again, see~\cite{pibiri2018inverted} for a detailed description of such encoding as well as of its \emph{partitioned} variant {\pef}~\cite{ottaviano2014partitioned}.
We test this encoder as (one of the) representative of the best space/time trade-off in the literature.
Not surprisingly, the adoption of this encoder to model the levels of a trie has been recently used to reduce the space of massive $n$-gram datasets by a factor of 2 over the most compact trie representation and to provide state-of-the-art time performance~\cite{pibiri2017efficient,Pibiri2019}.

Other approaches focus on representing a sequence of $d$-gaps, i.e., differences between successive integers, with the requirement of computing a prefix-sum when decoding the $d$-gapped stream.
A common way of speeding up the operations on the compressed sequence, is to divide the sequence into fixed-size blocks, e.g., 128 integers.
As a meaningful representative of such technique, we use Variable-Byte ({\vbyte})~\cite{thiel1972program} that achieves the highest decompression speed in the literature, especially when combined with the SIMD-ized decoding algorithm devised by Plaisance et al.~\cite{maskedvb}.
Variable-Byte compresses an integer using the minimum number of bytes needed
by its binary representation. The binary representation is divided into
chunks of 7 \emph{data} bits with an additional \emph{control} bit
inserted to indicate whether the byte is the last byte of the integer.
It follows that its simple and byte-aligned nature strongly favours
decoding efficiency at the expense of compression effectiveness.

\myparagraph{Performance}
Table~\ref{tab:compressors} reports the performance of the mentioned techniques  when applied to represent the node ID sequences of the tries of {\dbpedia}.
The space is expressed as bits per triple (henceforth, {\bpt}), whereas the efficiency of the operations {\access}, {\find} and {\scan} in nanoseconds per integer.
The reported timings resulted from the average across five runs of the same experiment to smooth fluctuations during measurements.

In order to operate in a meaningful benchmarking setting for the {\select} algorithm, instead of accessing random positions in the sequences or resolving a {\find} in random intervals, we use a set of 5,000 actual triples randomly extracted from the indexed {\dbpedia} dataset.
Given a triple $(s,p,o)$ belonging to the trie {\spo}, we report in the table the time required to \func{access} the second level at the position of $p$ (pre-calculated); similarly, we measure the time needed for finding the position of $p$ among the children of $s$ (\func{find}).
The time reported for \func{scan} indicates the time spent per node, when decoding the level sequentially.
The same holds for the third level of {\spo}, as well as for the other tries {\pos} and {\osp}.
The details of the used test machine are reported at the beginning of Section~\ref{sec:experiments}.
%Here we just mention that our code is written in C++14, compiled with gcc 7.3.0 using the highest optimization setting and tested on a machine equipped with Intel i7-7700 processors, 64 GB of RAM and running Linux 4.4.0, 64 bits.

%\return
Now, by looking at the results reported
in Table~\ref{tab:compressors}, we can express
the following considerations.

\begin{itemize}
\item All the tested techniques require about the same space for representing the sequences of subjects and objects, with {\pef} being the most space-efficient.

\item There is only a little difference between the space of {\pef} and {\comp} when representing the objects of {\spo} (third level).
The reason for this behavior is the \emph{monotonicity} of the sequences required by the Elias-Fano encoder.
In fact, while the sequences of pointers are monotone by construction and, thus, are immediately encodable, the same does not hold for node ID sequences where only sub-sequences of sibling nodes are ordered. Node ID sequences are thus formed by \emph{sorted sub-sequences} with the last element of a sub-sequence being not necessarily smaller than the first one of the next one. In order to encode such sequences with a compressor like Elias-Fano
we need to apply a simple  transformation adding to the value of each node ID the value of the prefix-sum of the previously coded sub-sequence.
This transformation makes Elias-Fano perform poorly
whenever the prefix-sum is updated too frequently as it happens in the presence of many small ranges, because the universe of representation grows (very) quickly, thus enlarging the space of representation.
As we will show later, most levels of the tries are indeed populated by such tiny ranges, meaning that each node has only few children on average.

\item {\pef} is roughly 2$\times$ smaller than the other options on the sequences of predicates.

\item The {\comp} representation is, in general, the fastest for all operations, especially when performing random {\access} that requires 1.4 -- 2.6 nanoseconds.
However, it is the least space-efficient as well.

\item The {\pef} variant imposes a penalty of roughly 2$\times$ over its un-partitioned counterpart when performing random {\access}, but results to be faster on average for the {\find} operation because it operates inside a partition that is much smaller than the whole sequence.
As expected, {\vbyte} codes spend more time for {\find} because of the (expensive) decompression of blocks that happens to be competitive only when decoding the sequences of predicates.
\end{itemize}

\return
\noindent
For the reasons discussed above, we adopt the following design choices.
We use {\pef} to represent all the sequences of node IDs, except for the last of the trie {\spo} where we adopt the {\comp} representation.
The pointer sequences are represented using plain {\ef} (not partitioned) given its superior random {\access} efficiency.

From now on, we refer to this solution as the {\three} index.

\myparagraph{Space breakdowns}
Lastly in this subsection, we discuss the space breakdowns reported in parentheses in Table~\ref{tab:compressors}. These values indicate the percentage of space out of the whole index required  to encode with a given compressor the specific sequence of IDs (the space for pointers is excluded from the count).

As a general consideration, we observe that each permutation takes roughly 1/3 of the space of the whole index, with the {\spo} trie being a little larger than the other two.
For example, by considering the values for the {\pef} encoder, we have that the levels of {\spo} require 2.51\% + 32.95\% = 35.46\% of the whole index, whereas the ones of {\pos} 27.06\%.
Summing up all the values for {\pef}, we get a total percentage of 91.05\%, thus the space for the pointer sequences takes less than 9\% of the whole index.

The levels of the index that contribute the most to the overall index space are: the third levels of {\spo} and {\pos}; the second level of {\osp}.
Specifically, we observe that such levels take practically the \emph{whole} space of the single tries which they belong to because the
contribution of the other two levels is marginal
%other two levels take just a small fraction of the space
(e.g., between 5\% and 7.5\% for {\pef}).
The first levels occupy almost no space because they comprise only the pointers.
The second levels of {\spo} and {\pos} have a small memory footprint
for different reasons: for the former, because the number of bits needed to represent a predicate is small compared to the number of bits necessary to represent subjects and objects (see Table~\ref{tab:datasets}); for the latter, because the high associativity of predicates partitions the objects' space into a few but very long sorted sequences that are highly compressible. For example, only 1.34 {\bpt} are needed for the objects to be represented in the second level of the trie {\pos}. The former reason also causes the third level of {\osp} to require a small space.

These considerations suggest that the efforts in trying to reduce the space of our RDF index should be mainly targeted  to reduce the space for encoding level three of {\spo} and {\pos} and level two of {\osp}. This will be the objective of the next section.

\subsection{Cross compression}\label{subsec:remapping}
The index layout that we have introduced represents the triples three times in different (cyclic) permutations in order to optimally solve all triple selection patterns.
However, the current description does not take advantage of the fact that in this way the same set of triples is represented multiple times and that, consequently, the index has an abundance of redundant information.
In this section we develop a novel compression framework that does not compress each trie independently but  employs levels of the tries to compress levels of other tries, thus holistically \emph{cross-compressing} the different permutations.

\begin{figure}[t]
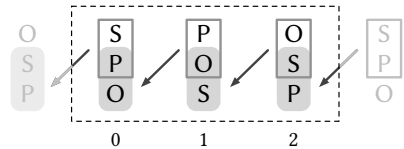

\centering
\includegraphics[scale=0.45]{{{imgs/remapping}}}
\caption{Graphical representation of our cross-compression technique, in which an arrow $(\textsf{X})_i \rightarrow (\textsf{Y})_j$ indicates that level \textsf{X} of trie $i$ is used to cross-compress the level \textsf{Y} of trie $j = (i + 2)$ mod $3$, for $i = 0, 1, 2$.}
\label{fig:remapping}
\end{figure}

The compression framework is graphically illustrated in Fig.~\ref{fig:remapping}. In the picture, we depict the three different permutations and highlight that the levels enclosed in the squared boxes are used to cross-compress the (lower) levels enclosed by the shaded boxes as pointed to by the arrows.
Cross-compression works by noting this crucial property:
\emph{the nodes belonging to a sub-tree rooted in the second level of trie $j$ are a subset of the nodes belonging to a sub-tree rooted in the first level of trie $i$, with} $j = (i + 2)$ mod $3$\emph{, for $i = 0, 1, 2$.}
The correctness of this property follows automatically by taking into account that the triples indexed by each permutation are the same.
Therefore, the children of $x$ in the second level of trie $j$ can be re-written as the \emph{positions} they take in the (larger, enclosing) set of children of $x$ in the first level of trie $i$.
Re-writing the node IDs as positions relative to the set of children of a sub-tree yields a clear space optimization because the number of children of a given node is much smaller (on average) than the number of distinct subjects or objects. (See also Table~\ref{tab:datasets} for the precise statistics.)

\begin{table}[t]
\centering
\caption{Number of children of the trie nodes for {\dbpedia}.}
\scalebox{\mytablescale}{
    \begin{tabular}{
c
c
r
r
}

\toprule

Trie
& Level
& Average
& Maximum
\\
%& \multicolumn{1}{c}{\dblp}
%&
%& \multicolumn{1}{c}{\gnames}
%&
%& \multicolumn{1}{c}{\dbpedia} \\

%\cmidrule(lr){3-3}
%\cmidrule(lr){5-5}
%\cmidrule(lr){7-7}

\midrule

\multirow{2}{*}{{\spo}}
& {1}
%& 11.41
%%& 18
%&
%& 14.19
%%& 24
%&
& 5.54
& 52
\\

& {2}
%& 1.51
%%& 617
%&
%& 1.03
%%& 276
%&
& 2.32
& {8,489}
\\

\midrule

\multirow{2}{*}{{\pos}}
& {1}
%& 1720286.44
%%& 10033344
%&
%& 1715219.85
%%& 8345450
%&
& {91,578.32}
& {21,219,244}
\\

& {2}
%& 1.90
%%& 3364083
%&
%& 2.75
%%& 8345450
%&
& 2.59
& {10,141,311}
\\

\midrule

\multirow{2}{*}{{\osp}}
& {1}
%& 1.93
%%& 3364083
%&
%& 2.68
%%& 8345450
%&
& 2.70
& {10,141,327}
\\

& {2}
%& 1.26
%%& 3
%&
%& 1.10
%%& 3
%&
& 1.13
& 10
\\

\bottomrule

\end{tabular}

}
\label{tab:range_stats}
\end{table}

We claim that the average number of children for the nodes in the first levels of the tries is the key statistic affecting the effectiveness of the described cross-compression technique because smaller numbers require a smaller number of bits to be represented.
We report this statistic in Table~\ref{tab:range_stats} for the {\dbpedia} dataset.
The data reported in the table is self-explanatory: except for the first level of {\pos}, the average number of children is \emph{very} low, being actually less than 3 for {\osp}, across the different permutations.
As already mentioned, the high associativity of the predicates causes each predicate to have many children, i.e., several orders of magnitude more than the children of subjects and objects.
We also show the maximum number of children, i.e., a maximum of $m$ indicates that there is a least one node that has $m$ children. While this value can be very distant from the average, for some levels like the second of {\spo} and {\osp} is actually similar, meaning that the selection patterns \textsf{SP?} and \textsf{S?O} have a robust worst-case guarantee.

Fig.~\ref{alg:map-unmap} shows a compact pseudo code illustrating how a \var{child} ID can be rewritten conditionally to its \var{parent} ID in the described cross-compression framework.
Conversely, the function \func{unmap} shows what actions are needed to recover the original ID before returning the result to the user. In particular, this latter function needs to be applied to the third component of \emph{every} triple matched by a selection pattern, hence partially eroding the retrieval efficiency.
Note that fast random access is needed by the \func{unmap} function, thus it could be convenient to model \var{levels}$[1]$.\var{nodes} with a {\comp} representation (see also Table~\ref{tab:compressors} for benchmark numbers).

\myparagraph{Discussion}
We have seen that the nodes in the third level of the {\pos} trie can be cross-compressed using the sub-trees rooted in the first level of the trie {\osp}, whose average number of children is actually less than 3 on real-world datasets.
Therefore, what we should expect is to have the third level of {\pos} mostly populated with values in $\{$0, 1, 2$\}$ that requires just two bits to be written instead of more than 20 bits to represent each subject.
Therefore, we argue that a significant space saving can be achieved by applying this technique, introducing a slowdown for two (out of eight) selection patterns only, i.e., \textsf{?PO} and \textsf{?P?} that are solved on the trie {\pos}.
%(see Section~\ref{sec:experiments} for a detailed experimental analysis).
As already motivated, the second level of {\osp} is represented with {\comp}.

We also argue that cross-compressing the other two permutations, i.e., {\spo} and {\osp}, yields only modest advantages. Again, refer to Table~\ref{tab:range_stats}. To cross-compress the third level of {\spo} we use the children branching out from the predicates to the objects. But, as already noted, a predicate has many children, thus dwarfing the corresponding reduction in the average number of bits needed to represent a mapped ID.
For the permutation {\osp}, instead, we have to shrink the node ID of the predicates. But these are already encoded compactly, e.g., in less than 4.8 {\bpt} on average for {\dbpedia} (see Table~\ref{tab:compressors}).
Thus, given that we can not expect great space savings in these two cases, we consider the cross-compressed index
to be the one with cross-compression on the {\pos} triples
and that we indicate with {\threecc} in the following.

\begin{figure}[t]
\removelatexerror
\scalebox{\mytablescale}{\input{pos-map.tex}}
\scalebox{\mytablescale}{\input{pos-unmap.tex}}
\caption{Functions used to \func{map} and \func{unmap} a child ID \emph{conditionally} to its sibling IDs.
\label{alg:map-unmap}}
\end{figure}

%Before concluding the section, we discuss an alternative approach to cross-compress the third level of {\spo}.
%Instead of using the level \textsf{PO} (as illustrated in Fig.~\ref{fig:remapping}), we could map an object \var{o} in the triple (\var{s}, \var{p}, \var{o}) as the position it takes in the sequence of all the distinct, sorted, objects that appear in the triples having subject \var{s}.
%In other words, we could use the level \textsf{SO} to cross-compress the third level of {\spo}.
%Representing the second level of the trie {\osp} with a wavelet tree~\cite{grossi2003} allows us to navigate the level \textsf{OS} backward, thus implementing the functions in Fig.~\ref{alg:map-unmap} using binary rank/select primitives.
%We tested this approach using the (integer-valued) wavelet tree implementation from the
%\emph{SDSL} library~\cite{gbmp2014sea} and achieved further reductions in space as expected.
%However, the large alphabet of representation (equal to the number of subjects) causes the wavelet tree to be tall, e.g., more than 20 levels, hence the non-cache-friendly traversal of these many levels had a disrupting effect on the performance values we reported in Table~\ref{tab:compressors}, by even requiring microseconds per integer.
%Therefore, we regard this alternative to be not competitive with the solutions already described,
%especially regarding the efficient resolution of triple selection patterns.

\subsection{Eliminating a permutation}\label{subsec:twotries}
The low number of predicates exhibited by RDF data in combination to
the corresponding very few children of the nodes in the second level of the trie {\osp}, leads us to consider a different \func{select} algorithm for the resolution of the query pattern \textsf{S?O}, able to take advantage of such skewed distribution of the predicates.
In fact, recall from Table~\ref{tab:range_stats} that on the {\dbpedia} dataset, for a given (\var{subject}, \var{object}) pair, just 1.13 predicates are returned on average
and 10 \emph{at most}.
Also recall from Section~\ref{subsec:core} that the pattern \textsf{S?O} is solved
on the trie {\osp} trie
with the \func{\select} algorithm in Fig.~\ref{alg:select} by performing a \func{find} operation followed by a \func{scan}.

\myparagraph{An alternative approach for solving \textsf{S?O}}
The idea is to pattern match \textsf{S?O} directly over the {\spo} permutation with the algorithm illustrated in Fig.~\ref{alg:enumerate}.
For a given subject $s$ and object $o$, in short, we operate as follows.
We consider the set of all the predicates that are children of $s$.
For each predicate $p_i$ in the set, we determine if the object $o$ is a child of $p_i$ with the \func{find} function: if it is, then ($s$, $p_i$, $o$) is a triple to return.
We refer to such strategy as the \func{enumerate} algorithm, whose correctness is immediate.

As partially motivated, we argue that the efficiency of the outlined algorithm is due to several distinct facts.
\begin{itemize}
\item Although the worst-case time complexity of the algorithm is $O(1 + |\textsf{P}|(1 + \log|\textsf{O}|))$,
the small number of predicates as children of a given subject implies that the \textsf{for} loop in Fig.~\ref{alg:enumerate} performs few iterations.
While these children are few per se, e.g., \emph{at most} 52 for {\dbpedia}, the
iterations will be on average far less. For example, less than 6 iterations are needed on average for {\dbpedia}.
(See detailed statistics in Table~\ref{tab:range_stats}.)

\item For a given (subject, predicate) pair, we have a very limited number of children to be searched for $o$, thus making the \func{find} function run \emph{faster} by \emph{short scans} rather than via binary search. On {\dbpedia}, an average of 2.32 children have to be scanned per pair.

\item Furthermore, the \func{enumerate} algorithm operates on the {\spo} permutation whereas the \func{select} algorithm on {\osp}: but if we consider the rows for the {\pef} encoder in Table~\ref{tab:compressors},
we see that a \func{find} operation issued on the second level of {\osp} costs
1.6 times more the amount of nanoseconds spent on {\spo}.
Thus, avoiding percolating the trie {\osp} produces further savings.
\end{itemize}

\begin{figure}[t]
\removelatexerror
\scalebox{\mytablescale}{
    \input{enumerate.tex}
}
\caption{The algorithm specialized to pattern match \textsf{S?O}.
\label{alg:enumerate}}
\end{figure}

\begin{table*}
\centering
\caption{Datasets basic statistics, reporting the total number of triples,
number of distinct subjects (\textsf{S}),{\protect\\}distinct predicates (\textsf{P}), distinct objects (\textsf{O}) and distinct pairs.}
\scalebox{\mytablescale}{
    \begin{tabular}{lrrrrrrr}
\toprule
Dataset
    & \multicolumn{1}{c}{Triples}
        & \multicolumn{1}{c}{\textsf{S}}
        & \multicolumn{1}{c}{\textsf{P}}
        & \multicolumn{1}{c}{\textsf{O}}
        & \multicolumn{1}{c}{\textsf{SP} pairs}
        & \multicolumn{1}{c}{\textsf{PO} pairs}
        & \multicolumn{1}{c}{\textsf{OS} pairs}
\\

\midrule

\dblp
    & {88,150,324}
        & {5,125,936}
            & \pp\pp\pp\pp{27}
            & {36,413,780}
                    & {58,476,283}
                        & {46,468,249}
                            & {70,234,083}
\\
\gnames
    & {123,020,821}
        & {8,345,450}
            & \pp\pp\pp\pp{26}
            & {42,728,317}
                    & {118,410,418}
                        & {45,096,877}
                            & {112,961,698}
\\
\dbpedia
    & {351,592,624}
        & {27,318,781}
            & \pp\pp{1,480}
            & {115,872,941}
                    & {151,464,424}
                        & {135,673,814}
                            & {311,567,728}
\\
\wat
    & {1,092,155,948}
        & {52,120,385}
            & \pp\pp\pp\pp{86}
            & {92,220,397}
                    & {230,085,646}
                        & {111,561,465}
                            & {1,092,137,931}
\\
\lubm
    & {1,334,681,190}
        & {217,006,852}
            & \pp\pp\pp\pp{17}
            & {161,413,040}
                    & {1,060,824,925}
                        & {195,085,216}
                            & {1,334,459,593}
\\
\freebase
    & {2,067,068,154}
        & {102,001,451}
            & \hspace{-0.07cm}{770,415}
            & {438,832,462}
                    & {878,472,435}
                        & {722,280,094}
                            & {1,765,877,943}
\\
\bottomrule

\end{tabular}

}
\label{tab:datasets}
\end{table*}

\myparagraph{Discussion}
In the light of devising a competitive algorithm to pattern match \textsf{S?O} on {\spo}, we consider another index layout.
Combining the design introduced in Section~\ref{subsec:core} with the present considerations, \emph{five} out of the eight different selection patterns can be solved efficiently by the trie {\spo}, i.e.: \textsf{SPO}, \textsf{SP?}, \textsf{S??}, \textsf{S?O} and \textsf{???}.
In order to support two more selection patterns, we can either chose to:
(1) materialize the permutation {\pos} for \emph{predicate-based} retrieval (query patterns \textsf{?PO} and \textsf{?P?});
(2) materialize the permutation {\ops} for \emph{object-based} retrieval (query patterns \textsf{?PO} and \textsf{??O}).
The choice of which permutation to maintain depends on the statistics of the selection patterns that have to be supported.
%,as we are going to discuss in Section~\ref{subsec:overall}.
We stress that the introduced \func{enumerate} algorithm allows us to actually \emph{save the space} for a third permutation that, as we have already pointed out in Section~\ref{subsec:core}, costs roughly 1/3 of the whole space of the index.
We call this solution the {\two} index, with two concrete instantiations:
{\twop} (predicate-based) and {\twoo} (object-based).

The only selection pattern that is not immediately supported is \textsf{?P?} for {\twoo} (symmetrically, \textsf{??O} for {\twop}).
However, we know that predicates are very associative, thus the list of subject nodes having a predicate parent is, on average, very long and highly compressible.
Therefore, we can afford to maintain a two-level structure \textsf{PS} that, for every predicate $p$, maintains the list of all subjects that appear in triples having predicate $p$.
Recall from our discussion concerning the space breakdowns in Section~\ref{subsec:core} that the similar structure \textsf{PO} costs just 1.34 {\bpt} for {\dbpedia}, thus \textsf{PS} should be only a little more costly given that the \textsf{SP} pairs are more in number than the \textsf{PO} pairs (see Table~\ref{tab:datasets}).
Therefore, we proceed as follows.

\begin{enumerate}
\item \emph{Access the list $p \rightarrow [s_1,\ldots,s_n]$.}
\item \emph{For $i = 1,\ldots,n$, consider the pair $(s_i,p)$ and pattern match $s_i p ?$ against the permutation } {\spo}\emph{.}
\end{enumerate}

\noindent
Similarly for the index {\twop}, if we consider a query pattern \textsf{??O} with specified object $o$,
$|\textsf{P}|$ \func{find} operations are needed to locate all the occurrences of $o$ in the second level of the trie {\pos}.

\begin{enumerate}
\item \emph{For every predicate $p$, consider its children.}
\item \emph{Determine if $o$ is among them with the} \func{find} \emph{function. If yes, then return all triples} ($s, p, o$) \emph{with $s$ being a child of the pair $(p,o)$.}
\end{enumerate}

\noindent
Again, the correctness of these two algorithms is immediate.
Their worst-case time complexities are respectively $O(|\textsf{S}|(1 + \log|\textsf{P}|) + \occ)$ for \textsf{?P?} and $O(|\textsf{P}|(1 + \log|\textsf{O}|) + \occ)$
for \textsf{??O}.
In order to distinguish them from
\func{select} and \func{enumerate}, in the following analysis we refer to both approaches as \func{inverted}.

\section{Experimental Analysis}\label{sec:experiments}
In this section we first compare the performance of the three solutions that we have introduced in the previous section, i.e., {\three} (Section~\ref{subsec:core}), {\three} with \emph{cross compression}, indicated as {\threecc} (Section~\ref{subsec:remapping}) and {\two} (Section~\ref{subsec:twotries}).
Then, we compare against the competitive approaches at the state-of-the-art that we have discussed in Section~\ref{sec:related}.

\myparagraph{Datasets}
We perform our experimental analysis on the following standard datasets, whose statistics are summarized in Table~\ref{tab:datasets}.
%\begin{itemize}
%\item
{\dblp}~\cite{dblp} is the dump of the DBLP Computer Science Bibliography collected during 2017.
%and available for download at \url{http://www.rdfhdt.org/datasets}.
%\item
{\gnames}~\cite{geo} is the official 2012 dump collected by geonames.org.
%and available for download at \url{http://www.rdfhdt.org/datasets}.
%\item
{\dbpedia}~\cite{auer2007dbpedia} is the English version 3.9 of DBpedia,
dubbed as ``the nucleus for a Web of Data''.
%, and available for download at \url{http://www.rdfhdt.org/datasets}.
%\item {\btc} is the Billion Triples Challenge 2012 Dataset crawled during May/June 2012. The official page of the dataset is \url{https://km.aifb.kit.edu/projects/btc-2012/}.
%\item
{\wat}~\cite{watdiv} is the ``1B Triples'' Waterloo SPARQL Diversity Test Suite
dataset, available for download at \url{https://dsg.uwaterloo.ca/watdiv}
(Section 4).
%\giulio{SPARQL queries at \url{http://grid.hust.edu.cn/triplebit/watdiv.txt}}
%\item
{\lubm}~\cite{lubm} is a synthetic dataset generated using the methodology described by Lehigh University Benchmark,
that is an established and widely-used benchmark for semantic data.
(In particular, the dataset was generated using the tools from \url{https://github.com/rvesse/lubm-uba} with option \texttt{-u 10000}.)
%\giulio{SPARQL queries at \url{http://grid.hust.edu.cn/triplebit/lubm.txt}}
%\item
{\freebase}~\cite{freebase} is the last available data dump of Google Freebase collected in December 2013.
%\end{itemize}

\myparagraph{Experimental setting and methodology}
All the experiments are performed on a server machine with 4 Intel i7-7700 cores (@3.6 GHz), with 64 GB of RAM DDR3 (@2.133 GHz), running Linux 4.4.0, 64 bits.
We implemented the indexes in C++14 and compiled the code with gcc 7.3.0 using the highest optimization setting, i.e., with compilation flags \texttt{-O3} and \texttt{-march=native}.

The indexes are saved to the disk after construction and loaded in internal memory to be queried.
In order to avoid disk caching effects, we clear the cache of the disk before measuring query timings.
To measure the query processing speed, we use a set of 5,000 triples
drawn at random from the datasets and set 0, 1 or 2 wildcard symbols.
%Clearly, 3 wildcards correspond to a scan of the index that returns all the stored triples.
The reported results are averages over five runs of queries in order to smooth fluctuations during measurements.
All queries run on a single processing core.

In all tables, speed up factors are taken
with respect to the values highlighted in bold font.
Notation $x$ (+$p$\%) means that if we subtract $p$\% of the space of $x$,
we obtain the value in bold font, say \textbf{\emph{z}}.
In other words:
$x$ takes $p$\% more space than \textbf{\emph{z}}.

\subsection{The permuted trie index}
To assess the performance of our solutions we chose to show the results for the \emph{real-world} datasets of {\dblp}, {\gnames}, {\dbpedia} and {\freebase}, given that similar trends and conclusions were observed on the others.
We will use the other two (synthetic) datasets, {\wat} and {\lubm}, to test the speed of selection patterns occurring in the corresponding SPARQL query logs in Section~\ref{subsec:overall}.

\myparagraph{Compression effectiveness}
We first comment on the compression effectiveness achieved by our different solutions.
Refer to the upper part of Table~\ref{tab:comparison}, where we report the average number
of bits spent per represented triple on the different datasets.
We sorted the rows in order of increasing effectiveness, thus we can immediately conclude
that the most compact index is {\twop}, with {\threecc} in between the values of {\three} and {\two}.
The {\twop} variant is even smaller than {\twoo} because it materializes the permutation {\pos} that
requires less space than {\ops} as maintained by {\twoo}.
In particular, the compression level exhibited by {\three} lies in the range of 72 -- 81 {\bpt}
and as expected, the refinements devised in Section~\ref{subsec:remapping} and~\ref{subsec:twotries}
pay off.
The {\threecc} index brings a further saving of 11\% on average, because
the space of representation of the third level of {\pos} is almost dissolved, being actually reduced
by roughly 4$\times$ on the different datasets (but the {\comp} compressor on the second level of {\osp}
takes more space than {\pef}, as used by {\three}).
Instead, the {\two} indexes reduces the space by 30\% on average since they basically
eliminate one permutation of the triples, hence optimizing by roughly 1/3.

\myparagraph{Triple selection patterns}
Now we examine the speed of all the different triple selection patterns by commenting on the lower
part of Table~\ref{tab:comparison}, where we report the timings in nanoseconds spent per returned triple.
In particular, patterns are grouped together when solved by the same trie (or have exactly the
same performance).
Let us proceed in order, by discussing each group.

\begin{figure*}[t]
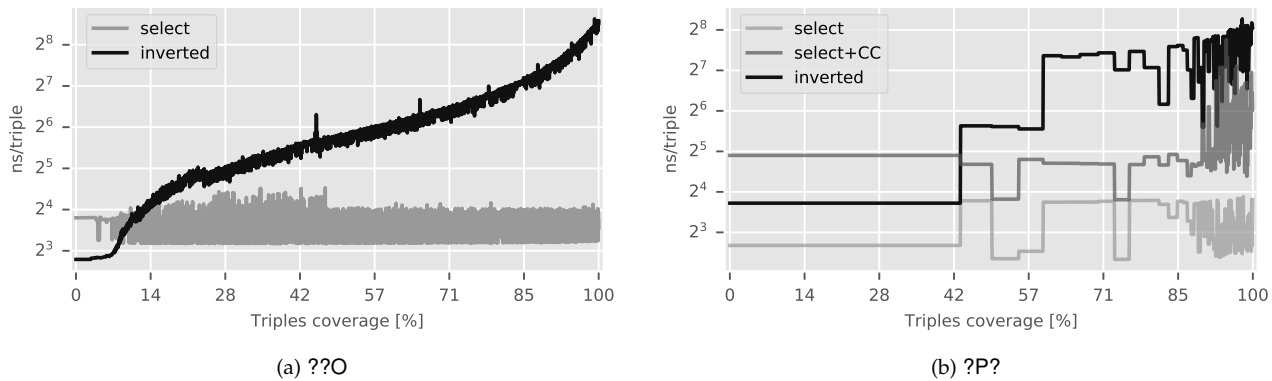

\centering

\subfloat[\textsf{??O}]{
	\includegraphics[scale=0.9]{{{plots/objects2}}}
	\label{fig:objects}
}
\subfloat[\textsf{?P?}]{
	\includegraphics[scale=0.9]{{{plots/predicates2}}}
	\label{fig:predicates}
}

\caption{Average {\nt} by \emph{decreasing} number of matches, for the query pattern \textsf{??O} and \textsf{?P?}.}
\end{figure*}

\begin{table}
\centering
\caption{Comparison between the performance of {\three}, {\threecc} and {\two} indexes, expressed as the total space in {\bpt}{\protect\\}and in average {\nt} for all the different selection patterns.
}
\scalebox{0.85}{
	\hspace{-0.3cm}\npdecimalsign{.}
\nprounddigits{2}
\begin{tabular}{
c@{\hspace{1pt}}
c@{\hspace{1pt}}
c@{\hspace{10pt}}
l@{\hspace{1pt}}
c@{\hspace{5pt}}
r@{\hspace{1pt}}
r@{\hspace{3pt}}
r
r@{\hspace{1pt}}
r@{\hspace{3pt}}
r
r@{\hspace{1pt}}
r@{\hspace{3pt}}
r
r@{\hspace{1pt}}
r@{\hspace{3pt}}
}

\toprule

\multicolumn{3}{c}{}
& Index
&
& \multicolumn{2}{c}{\dblp}
&
& \multicolumn{2}{c}{\gnames}
&
& \multicolumn{2}{c}{\dbpedia}
&
& \multicolumn{2}{c}{\freebase}
\\

\midrule

&&&
&
& \multicolumn{2}{c}{\bpt}
&
& \multicolumn{2}{c}{\bpt}
&
& \multicolumn{2}{c}{\bpt}
&
& \multicolumn{2}{c}{\bpt}
\\

\cmidrule(lr){6-7}
\cmidrule(lr){9-10}
\cmidrule(lr){12-13}
\cmidrule(lr){15-16}

&&&
{\three}
&
& 75.24 & {(+31\%)}
&
& 71.59 & {(+32\%)}
&
& 80.64 & {(+33\%)}
&
& 74.20 & {(+30\%)}
\\

&&&
{\threecc}
&
& 63.54 & {(+18\%)} % 58.90
&
& 67.04 & {(+27\%)} % 59.33
&
& 66.91 & {(+19\%)} % 62.82
&
& 70.46 & {(+26\%)} % 58.15
\\

&&&
{\twoo}
&
& 56.46 & {(+8\%)}
&
& 53.23 & {(+8\%)}
&
& 57.51 & {(+6\%)}
&
& 55.72 & {(+6\%)}
\\

&&&
{\twop}
&
& \bb{51.99} &
&
& \bb{48.98} &
&
& \bb{54.14} &
&
& \bb{52.17} &
\\

\midrule

&&&
&
& \multicolumn{2}{c}{\nt}
&
& \multicolumn{2}{c}{\nt}
&
& \multicolumn{2}{c}{\nt}
&
& \multicolumn{2}{c}{\nt}
\\

\cmidrule(lr){6-7}
\cmidrule(lr){9-10}
\cmidrule(lr){12-13}
\cmidrule(lr){15-16}

$\textsf{S}$ & $\textsf{P}$ & $\textsf{O}$
& \emph{all} % {\three},{\threecc},{\twox}
&
& \pp 203 &
&
& \pp 221 &
&
& \pp 353 &
&
& \pp 521 &
\\

$\textsf{S}$ & $\textsf{P}$ & $?$
& \emph{all} % {\three},{\threecc},{\twox}
&
& \pp 197 &
&
& \pp 347 &
&
& \pp\pp 11 &
&
& \pp\pp\pp 3 &
\\

$\textsf{S}$ & $?$ & $?$
& \emph{all} % {\three},{\threecc},{\twox}
&
& \pp\pp 28 &
&
& \pp\pp 40 &
&
& \pp\pp 10 &
&
& \pp\pp\pp 3 &
\\

$?$ & $?$ & $?$
& \emph{all} % {\three},{\threecc},{\twox}
&
& \pp\pp 11 &
&
& \pp\pp 13 &
&
& \pp\pp\pp 9 &
&
& \pp\pp\pp 9 &
\\ 

\midrule

\multirow{2}{*}{$\textsf{S}$}
&
\multirow{2}{*}{$?$}
&
\multirow{2}{*}{$\textsf{O}$}

& {\three},{\threecc}
&
& 2,490 & (5.6$\times$)
&
& 3,767 & (7.7$\times$)
&
& 1,833 & (2.6$\times$)
&
& 6,547 & (1.8$\times$)
\\

&&
& {\twoo},{\twop}
&
& \pp \textbf{445} &
&
& \pp \textbf{490} &
&
& \pp \textbf{692} &
&
&     \textbf{3,736} &
\\
\midrule

\multirow{2}{*}{$?$}
&
\multirow{2}{*}{$\textsf{P}$}
&
\multirow{2}{*}{$\textsf{O}$}

& {\three},{\twoo},{\twop}
&
& \pp\pp\pp \textbf{5} &
&
& \pp\pp\pp \textbf{5} &
&
& \pp\pp\pp \textbf{5} &
&
& \pp\pp\pp \textbf{5} &
\\

&&
& {\threecc}
&
& \pp\pp 12 & (2.4$\times$) % \pp\pp 74
&
& \pp\pp 15 & (3.0$\times$) % \pp\pp 80
&
& \pp\pp 16 & (3.2$\times$) % \pp\pp 78
&
& \pp\pp 14 & (2.8$\times$) % \pp\pp 80
\\
\midrule

\multirow{3}{*}{$?$}
&
\multirow{3}{*}{$?$}
&
\multirow{3}{*}{$\textsf{O}$}

& {\three},{\threecc}
&
& \pp\pp 12 &  (2.4$\times$)
&
& \pp\pp 12 &  (2.4$\times$)
&
& \pp\pp 12 &  (2.4$\times$)
&
& \pp\pp 10 &  (2.0$\times$)
\\

&&
& {\twoo}
&
& \pp\pp\pp \textbf{5} &
&
& \pp\pp\pp \textbf{5} &
&
& \pp\pp\pp \textbf{5} &
&
& \pp\pp\pp \textbf{5} &
\\

&&
& {\twop}
&
& \pp\pp\pp 5 &  (1.0$\times$)
&
& \pp\pp\pp 5 &  (1.0$\times$)
&
& \pp\pp\pp 6 &  (1.2$\times$)
&
& \pp\pp   10 &  (2.0$\times$)
\\

\midrule

\multirow{3}{*}{$?$}
&
\multirow{3}{*}{$\textsf{P}$}
&
\multirow{3}{*}{$?$}

& {\three},{\twop}
&
& \pp\pp\pp \textbf{9} &
&
& \pp\pp\pp \textbf{8} &
&
& \pp\pp\pp \textbf{6} &
&
& \pp\pp\pp \textbf{6} &
\\

&&
& {\threecc}
&
& \pp\pp 21 & (2.3$\times$) % \pp\pp 46
&
& \pp\pp 36 & (4.5$\times$) % \pp\pp 71
&
& \pp\pp 30 & (5.0$\times$) % \pp\pp 72
&
& \pp\pp 29 & (4.8$\times$) % \pp\pp 93
\\

&&
& {\twoo}
&
& 81 & (9.0$\times$)
&
& 138 & (17.2$\times$)
&
& 22 &  (3.7$\times$)
&
& 18 & (3.0$\times$)
\\

\bottomrule
\end{tabular}

}
\label{tab:comparison}
\end{table}

\begin{itemize}

\item The index component represented by the trie {\spo} is common by the three solutions, thus the results
for \textsf{SPO}, \textsf{SP?}, \textsf{S??}, \textsf{???} are the same.
We can see that the indexes solve a lookup operation, i.e., the query pattern \textsf{SPO} with all symbols
specified, in a fraction of a microsecond, specifically 1/5 -- 1/2 $\mu$sec.
This is basically the time needed for two \func{find} operations, respectively on the second and third levels of the trie.
Observe how, instead, the ``fixed'' cost of the \func{find} operation on the second level is amortized
by the number of matching triples for the pattern \textsf{SP?} on the larger {\dbpedia} and {\freebase}.
Similarly, the time spent per triple by the patterns \textsf{S??} and \textsf{???} is just a handful of nanoseconds.

\item We now discuss the pattern \textsf{S?O} that is solved by {\three} and {\threecc} with the \func{select} algorithm
(Fig.~\ref{alg:select}) on the trie {\osp}, but by {\two} with the \func{enumerate} algorithm
(Fig.~\ref{alg:enumerate}) on the trie {\spo}.
In Section~\ref{subsec:twotries}, we have discussed about the efficiency of the \func{enumerate} algorithm and motivated
its advantages over the \func{select} algorithm.
In fact, we observe from Table~\ref{tab:comparison} that \func{enumerate} is faster than
\func{select} by 4.4$\times$ on average (a minimum of 1.8$\times$ on {\freebase}, and up to 7.7$\times$ on {\gnames}).
In particular, we have claimed that the crucial statistic affecting the performance of the \func{enumerate} algorithm
is the very low associativity of the subjects, i.e., the limited number of predicates as children of a given subject.

\begin{figure}[t]
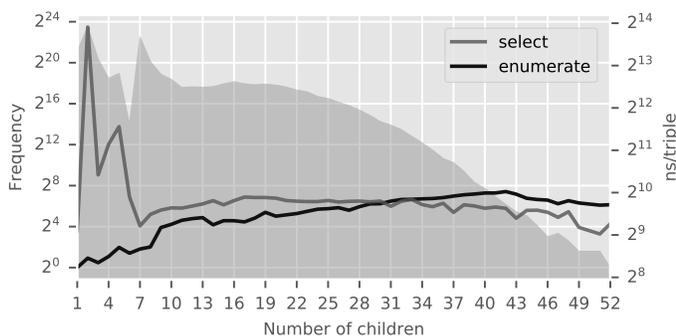

\includegraphics[scale=0.9]{{{plots/select_vs_enumerate}}}
\caption{Comparison between the \func{select} and \func{enumerate} algorithm when pattern matching \textsf{S?O} on {\dbpedia}, for queries having subjects whose number of children (predicates) is between 1 and 52. On the background, we show the distribution of the number of children.}
\label{fig:select_vs_enumerate}
\end{figure}

Therefore, in order to validate our hypothesis, we inspect the full spectrum of possibilities in Fig.~\ref{fig:select_vs_enumerate}, where we illustrate how the time for \textsf{S?O} changes for queries
whose subjects have different number of children.
We show the result on {\dbpedia} for space constraints (similar behaviors were observed for the other datasets anyway).
For {\dbpedia} the number of children of a subject, henceforth indicated with $C$, varies between 1 and 52 (see also Table~\ref{tab:range_stats}).
As it is clear from the plot, whenever $C < 30$ the \func{enumerate} algorithm is faster and looses only for higher
values, because of the cache-friendly behaviour of \func{select} due to its \func{scan}-based nature.
In particular, on the background of the plot we also show the distribution of $C$, i.e., how many subjects have $C = c$
children, for $c = 1..52$.
The distribution explains that the majority of subjects have only few children, and in
correspondence of such values, the \func{enumerate} algorithm is \emph{much} faster than \func{select} (e.g., up to
43$\times$ for $C=2$), thus explaining why on average is faster as well.

\item The pattern \textsf{?PO} illustrates how the cross-compression affects the retrieval of the subjects on the third level of the {\pos} permutation. For every match, the \func{unmap} function must be executed, hence requiring a random \func{access} to be performed on the second level of {\osp} (a potential cache-miss). This results in a slowdown of roughly 3$\times$.

\item The number of nanoseconds per triple for the pattern \textsf{??O} is controlled by the average selectivity of the queries:
while most of the queries have only a few matches, the presence of low-selective queries maintains the
average response time very low.
Also, notice that it is solved 2.4$\times$ faster by {\twoo} because it traverses the trie {\ops} and not {\osp} as it is done by the {\three} and {\threecc} indexes, thus employing \func{find} operations on the sequence of predicates rather than subjects that are faster (see details in Table~\ref{tab:compressors}).
It is also interesting to note that the pattern is solved efficiently by the {\twop} variant on average,
given that the cost of the $|\textsf{P}|$ \func{find} operations is well amortized by the number of returned matches.
However, when $|\textsf{P}|$ grows significantly and the results per query are very few,
the response time per occurrence is much higher. This \emph{output-sensitive} behaviour is a well-known
characteristic of RDF stores.
Therefore, for completeness, we also show in Fig.~\ref{fig:objects} how the time per triple changes by \emph{decreasing} number of matches, until we cover the whole set of triples.
The experiment shows that while the \func{select} is less output-sensitive, indeed for a reasonably good fraction of the
triples, e.g., 25\%, the \emph{average} query time of \func{inverted} is close to the one of \func{select} and even faster
for $\approx$10\% of the triples.

\item Similar considerations hold for the \textsf{?P?} query pattern, whose corresponding
stress behaviour is shown in Fig.~\ref{fig:predicates}: the \func{select} offers
the best worst-case guarantee across all different output sizes, with the cross-compression
technique (\func{select+CC}) and the \func{inverted} algorithm both imposing overheads
due to additional cache-misses. Observe how \func{inverted} is actually faster than
\func{select+CC} for more than 40\% of the triples, with the latter being in trade-off
position between the former and \func{select}.
\end{itemize}

\myparagraph{Range queries}
To asses the efficiency of range queries, supported
as we explained in Section~\ref{subsec:core},
we used the {\wat} dataset that contains several numeric
types (primarily integers and dates).
We tested query patterns of the form \textsf{?P?} and \textsf{?PO}
using the {\twop} index. Range constraints are expressed
on the object components, hence the queries are handled by the
trie {\pos} of {\twop}. 
We determined an average running time of 4.3 {\nt}.
The extra space of the data structure that we use to
support such queries
is very small as expected -- less than 0.1 {\bpt} --
when compressed using {\pef} as we did in our implementation.
In conclusion, range queries are supported efficiently
in both time and space regards.

\myparagraph{Discussion}
From the above considerations concerning the performance of our solutions,
we conclude that:
(1) the {\three} index is the one delivering best \emph{worst-case} performance guarantee
for all triple selection patterns;
(2) the {\two} variants reduce its space of representation by
25 -- 33\% \emph{without} affecting or even improving the retrieval efficiency on most triple
selection patterns (\emph{only one} pattern has a lower query throughput in the worst-case);
(3) the cross-compression technique is outperformed by the {\two} index layouts for space usage
but offers a better worst-case performance guarantee than {\twop} for the pattern \textsf{?P?}.
In particular, we have shown that \emph{only one} selection pattern out of the eight possible
has a lower query throughput in the worst-case for the {\two} indexes (\textsf{??O} on {\twop};
\textsf{?P?} on {\twoo}).

Therefore, as a reasonable trade-off between space and time concerns,
we elect {\twop} as the solution to compare against the state-of-the-art alternatives
in the following.

\subsection{Overall comparison}\label{subsec:overall}
In this section, we compare the performance of our selected solution {\twop} against the competitive approaches {\hdt}~\cite{HDT2} and {\triplebit}~\cite{TripleBit} described in Section~\ref{sec:related},
since these both outperform {\rdfx}~\cite{RDF3xjournal} and {\bitmat}~\cite{BitMat}.

In particular, the experimental analysis provided by the authors of {\hdt}~\cite{HDT2}
reports that {\rdfx} is larger than their own index
by 3.8$\times$, 4.6$\times$ and 3$\times$
on {\dblp}, {\gnames} and {\dbpedia} respectively, and also slower by a factor of 4 -- 8$\times$ on most
selection patterns ({\hdt} scores worse only for the pattern \textsf{?P?} because of the penalty
induced by representing predicate sequences with wavelet trees, as we are going to confirm next).
This is not surprising given that all the six triple permutations, plus aggregate indexes, are materialized by {\rdfx}
with a severe query processing overhead due to expensive I/O transfers.
Similar results are obtained by the authors of {\triplebit}~\cite{TripleBit}, whose experiments on {\lubm} show that {\rdfx} is 2$\times$ larger and 2 -- 14$\times$ slower. In the same paper
{\bitmat} is shown to be even larger than {\rdfx} and up to several orders of magnitude slower\footnote{To confirm this result, we built the {\bitmat} index on {\dbpedia}, with flag \texttt{COMPRESS\_FOLDED\_ARRAY} as suggested by the author Medha Atre, and measured a space occupancy of 483.72 {\bpt}.}.

For both {\hdt} and {\triplebit}, we (obviously) do not consider the space needed
for the string dictionaries, as well as the time needed for dictionary access/lookup.
We use the C++ libraries as provided by the corresponding authors and available at
\url{https://github.com/rdfhdt/hdt-cpp} and
\url{https://github.com/nitingupta910/TripleBit}, respectively.
Furthermore, both libraries are compiled with the same compiler we used for our own code
(gcc 7.3.0) and using the same optimization flags 
\texttt{-O3} and \texttt{-march=native}.

%In particular, to ensure a fair comparison against {\rdfx}, we only account for the space of the three permutations {\spo}, {\pos} and {\osp} because these are sufficient to support all selection patterns,
%and we load the index structure into memory in order to avoid expensive I/O transfers.

Table~\ref{tab:overall} reports the space of the indexes and the timings for the different selection patterns, but excluding (due to page limit) the ones for
\textsf{SPO} and \textsf{???}: our approach is anyway faster for both by at least a factor of 3$\times$ ({\triplebit} does not support the query pattern {\textsf{SPO}}).
Concerning the space, we see that the {\twop} index is significantly more compact, specifically by 30\% and almost 60\% compared to {\hdt} and {\triplebit} respectively, on average across all different datasets ({\triplebit} fails in building the index on {\freebase}).
Concerning the speed of triple selection patterns, most factors of improved efficiency range in the interval 2 -- 5$\times$ and, depending on the pattern examined, we report peaks of 26$\times$, 49$\times$, 81$\times$ or even 2,057$\times$.
%As already pointed out, the low selectivity of certain query patterns make byte-aligned codes such
%as {\vbyte} to score slightly better than {\twop} for their faster sequential decoding capabilities,
%such as, for example, {\rdfx} when solving {\textsf{?P?}}.

To further confirm the results, we execute all triple selection patterns needed to solve
the SPARQL queries from the publicly available query logs of {\wat}\footnote{\url{http://grid.hust.edu.cn/triplebit/watdiv.txt}}
and {\lubm}\footnote{\url{http://swat.cse.lehigh.edu/projects/lubm/queries-sparql.txt}}, that are both
well-established benchmarks for RDF data.
We used the query planning algorithm of {\triplebit} to obtain a serial decomposition of the
SPARQL queries into atomic selection patterns.
We argue that this methodology has the advantage of fairly testing the different
indexes over the same set of patterns as executed in the same order, thus really testing
the raw speed of the underlying indexing data structure
that is our goal in this work.
The results of the benchmark, illustrated in Table~\ref{tab:sparql}, closely match the ones
already discussed in Table~\ref{tab:overall}.
In particular, our index optimizes the space of representation by 21 -- 58\% and is
remarkably faster than {\triplebit} by 4.5$\times$ on average and by up to 238$\times$
than {\hdt} on {\wat}.
Most selection patterns for both query logs have \textsf{?P?} and \textsf{?PO} forms.
In fact, observe how our solution is
3 -- 6.7$\times$ faster than {\triplebit} on these patterns, with
{\hdt} being the worst because of the cache-misses introduced by representing the
predicate sequences with (potentially, tall) wavelet trees.

\begin{table}[t]
\centering
\caption{Comparison between the performance of different indexes, expressed as the total space in {\bpt} and in average {\nt}.}
\scalebox{0.8}{
	\hspace{-0.5cm}\npdecimalsign{.}
\nprounddigits{2}
\begin{tabular}{
c@{\hspace{1pt}}
c@{\hspace{1pt}}
c@{\hspace{10pt}}
l@{\hspace{1pt}}
c@{\hspace{5pt}}
r@{\hspace{1pt}}
r@{\hspace{3pt}}
r
r@{\hspace{1pt}}
r@{\hspace{3pt}}
r
r@{\hspace{1pt}}
r@{\hspace{3pt}}
r
r@{\hspace{1pt}}
r@{\hspace{3pt}}
}

\toprule

\multicolumn{3}{c}{}
& Index
&
& \multicolumn{2}{c}{\dblp}
&
& \multicolumn{2}{c}{\gnames}
&
& \multicolumn{2}{c}{\dbpedia}
&
& \multicolumn{2}{c}{\freebase}
\\

\midrule

&&&
&
& \multicolumn{2}{c}{\bpt}
&
& \multicolumn{2}{c}{\bpt}
&
& \multicolumn{2}{c}{\bpt}
&
& \multicolumn{2}{c}{\bpt}
\\

\cmidrule(lr){6-7}
\cmidrule(lr){9-10}
\cmidrule(lr){12-13}
\cmidrule(lr){15-16}

&&&
{\twop}
&
& \bb{51.99} &
&
& \bb{48.98} &
&
& \bb{54.14} &
&
& \bb{52.17} &
\\

&&&
{\hdt}
&
& 76.89 & {(+32\%)}
& 
& 88.73 & {(+45\%)}
&
& 76.66 & {(+29\%)}
&
& 83.11 & {(+37\%)}
\\

%&&&
%{\rdfx}
%&
%& 96.61 & {(+46\%)}
%&
%& 96.09 & {(+49\%)}
%&
%& 93.16 & {(+42\%)}
%&
%& 82.63 & {(+37\%)}
%\\

&&&
{\triplebit}
&
& 125.10 & {(+58\%)}
&
& 120.03 & {(+59\%)}
&
& 130.07 & {(+58\%)}
&
& {---} &
\\

\midrule

&&&
&
& \multicolumn{2}{c}{\nt}
&
& \multicolumn{2}{c}{\nt}
&
& \multicolumn{2}{c}{\nt}
&
& \multicolumn{2}{c}{\nt}
\\

\cmidrule(lr){6-7}
\cmidrule(lr){9-10}
\cmidrule(lr){12-13}
\cmidrule(lr){15-16}

%\multirow{3}{*}{\textsf{S}}
%&
%\multirow{3}{*}{\textsf{P}}
%&
%\multirow{3}{*}{\textsf{O}}
%
%& {\twop}
%&
%& \textbf{203} &
%&
%& \textbf{221} &
%&
%& \textbf{353} &
%&
%& \textbf{521} &
%\\
%
%&&
%& {\hdt}
%&
%& {1302} & (6.4$\times$)
%&
%& {1367} & (6.2$\times$)
%&
%& {1423} & (4.0$\times$)
%&
%& {1845} & (3.5$\times$)
%\\
%
%&&
%& {\rdfx}
%&
%& 928 & (4.6$\times$)
%&
%& 858 & (3.9$\times$)
%&
%& 1143 & (3.2$\times$)
%&
%& 2581 & (5.0$\times$)
%\\
%
%&&
%& {\triplebit}$^{\ast}$
%&
%& {---} &
%&
%& {---} &
%&
%& {---} &
%&
%& {---} &
%\\
%\midrule

\multirow{3}{*}{\textsf{?}}
&
\multirow{3}{*}{\textsf{P}}
&
\multirow{3}{*}{\textsf{O}}

& {\twop}
&
& \textbf{5} &
&
& \textbf{5} &
&
& \textbf{5} &
&
& \textbf{5} &
\\

&&
& {\hdt}
&
& {12} & (2.4$\times$)
&
& {13} & (2.6$\times$)
&
& {14} & (2.8$\times$)
&
& {13} & (2.6$\times$)
\\

%&&
%& {\rdfx}
%&
%& \textbf{3} & (0.6$\times$)
%&
%& \textbf{3} & (0.6$\times$)
%&
%& \textbf{3} & (0.6$\times$)
%&
%& \textbf{3} & (0.6$\times$)
%\\

&&
& {\triplebit}
&
& {15} & (3.0$\times$)
&
& {13} & (2.6$\times$)
&
& {14} & (2.8$\times$)
&
& {---}
\\
\midrule

\multirow{3}{*}{\textsf{S}}
&
\multirow{3}{*}{\textsf{?}}
&
\multirow{3}{*}{\textsf{O}}

& {\twop}
&
& \textbf{445} &
&
& \textbf{490} &
&
& \textbf{692} &
&
& \textbf{3736} &
\\

&&
& {\hdt}
&
& {1,789} & (4.0$\times$)
&
& {2,097} & (4.3$\times$)
&
& {3,010} & (4.3$\times$)
&
& {0.7$\times10^7$} & \footnotesize(2,057$\times$)
\\

%&&
%& {\rdfx}
%&
%& {3520} & (7.9$\times$)
%&
%& {5630} & (11.5$\times$)
%&
%& {3046} & (4.4$\times$)
%&
%& \hspace{-0.15cm}{10760} & (2.9$\times$)
%\\

&&
& {\triplebit}
&
& \hspace{-0.15cm}{11,872} & (26.7$\times$)
&
& \hspace{-0.15cm}{13,008} & (26.5$\times$)
&
& \hspace{-0.15cm}{18,023} & (26.0$\times$)
&
& {---} &
\\
\midrule

\multirow{3}{*}{\textsf{S}}
&
\multirow{3}{*}{\textsf{P}}
&
\multirow{3}{*}{\textsf{?}}

& {\twop}
&
& \textbf{197} &
&
& \textbf{347} &
&
& \textbf{11} &
&
& \textbf{3} &
\\

&&
& {\hdt}
&
& {640} & (3.2$\times$)
&
& {897} & (2.6$\times$)
&
& {30} & (2.7$\times$)
&
& {9} & (3.0$\times$)
\\

%&&
%& {\rdfx}
%&
%& {602} & (3.1$\times$)
%&
%& {956} & (2.8$\times$)
%&
%& {26} & (2.4$\times$)
%&
%& {3} & (1.0$\times$)
%\\

&&
& {\triplebit}
&
& {1,222} & (6.2$\times$)
&
& {927} & (2.7$\times$)
&
& {42} & (3.8$\times$)
&
& {---} &
\\
\midrule

\multirow{3}{*}{\textsf{S}}
&
\multirow{3}{*}{\textsf{?}}
&
\multirow{3}{*}{\textsf{?}}

& {\twop}
&
& \textbf{28} &
&
& \textbf{40} &
&
& \textbf{10} &
&
& \textbf{3} &
\\

&&
& {\hdt}
&
& {110} & (3.9$\times$)
&
& {154} & (3.9$\times$)
&
& {29} & (2.9$\times$)
&
& {9} & (3.0$\times$)
\\

%&&
%& {\rdfx}
%&
%& {61} & (2.2$\times$)
%&
%& {83} & (2.1$\times$)
%&
%& {20} & (2.0$\times$)
%&
%& {3} & (1.0$\times$)
%\\

&&
& {\triplebit}
&
& {2,275} & (81.2$\times$)
&
& {3,261} & (81.5$\times$)
&
& {490} & (49.0$\times$)
&
& {---} &
\\
\midrule

\multirow{3}{*}{\textsf{?}}
&
\multirow{3}{*}{\textsf{P}}
&
\multirow{3}{*}{\textsf{?}}

& {\twop}
&
& \textbf{9} &
&
& \textbf{8} &
&
& \textbf{6} &
&
& \textbf{4} &
\\

&&
& {\hdt}
&
& {108} & (12.0$\times$)
&
& {173} & (21.6$\times$)
&
& {32} & (5.3$\times$)
&
& {41} & (6.8$\times$)
\\

%&&
%& {\rdfx}
%&
%& 9 & (1.0$\times$)
%&
%& \textbf{7} & (0.9$\times$)
%&
%& 7 & (1.2$\times$)
%&
%& \textbf{4} & (0.7$\times$)
%\\

&&
& {\triplebit}
&
& {28} & (3.1$\times$)
&
& {28} & (3.5$\times$)
&
& {40} & (6.7$\times$)
&
& {---} &
\\
\midrule

\multirow{3}{*}{\textsf{?}}
&
\multirow{3}{*}{\textsf{?}}
&
\multirow{3}{*}{\textsf{O}}

& {\twop}
&
& \textbf{5} &
&
& \textbf{5} &
&
& \textbf{6} &
&
& \textbf{10} &
\\

&&
& {\hdt}
&
& {17} & (3.4$\times$)
&
& {17} & (3.4$\times$)
&
& {18} & (3.0$\times$)
&
& {18} & (1.8$\times$)
\\

%&&
%& {\rdfx}
%&
%& 9 & (1.8$\times$)
%&
%& 9 & (1.8$\times$)
%&
%& 9 & (1.5$\times$)
%&
%& \textbf{7} & (0.7$\times$)
%\\

&&
& {\triplebit}
&
& {24} & (4.8$\times$)
&
& {60} & (12.0$\times$)
&
& {24} & (4.0$\times$)
&
& {---} &
\\

%\midrule
%
%\multirow{3}{*}{\textsf{?}}
%&
%\multirow{3}{*}{\textsf{?}}
%&
%\multirow{3}{*}{\textsf{?}}
%
%& {\twop}
%&
%& 11 &
%&
%& 13 &
%&
%& \textbf{9} &
%&
%& \textbf{9} &
%\\
%
%&&
%& {\hdt}
%&
%& {52} & (4.7$\times$)
%&
%& {68} & (5.2$\times$)
%&
%& {40} & (4.4$\times$)
%&
%& {40} & (4.4$\times$)
%\\
%
%&&
%& {\rdfx}
%&
%& \textbf{10} & (0.9$\times$)
%&
%& \textbf{11} & (0.8$\times$)
%&
%& 9 & (1.0$\times$)
%&
%& 9 & (1.0$\times$)
%\\
%
%&&
%& {\triplebit}
%&
%& {15} & (1.4$\times$)
%&
%& {13} & (1.0$\times$)
%&
%& {63} & (7.0$\times$)
%&
%& {---} &
%\\

\bottomrule
\end{tabular}

}
\label{tab:overall}
\end{table}

\begin{table}[t]
\centering
\caption{Performance in {\bpt} and in average seconds spent by the different solutions to execute the sequence of triple selection patterns generated for the SPARQL queries in the {\wat} and {\lubm} logs.}
\scalebox{\mytablescale}{
	\hspace{-0.25cm}\npdecimalsign{.}
\nprounddigits{2}
\begin{tabular}{
l@{\hspace{1pt}}
c@{\hspace{5pt}}
r@{\hspace{1pt}}
r@{\hspace{5pt}}
r@{\hspace{1pt}}
r
c@{\hspace{5pt}}
r@{\hspace{1pt}}
r@{\hspace{5pt}}
r@{\hspace{1pt}}
r
}

\toprule
Index
&
& \multicolumn{4}{c}{\wat}
&
& \multicolumn{4}{c}{\lubm}
\\

\midrule

&
& \multicolumn{2}{c}{\bpt}
& \multicolumn{2}{c}{\sq}
&
& \multicolumn{2}{c}{\bpt}
& \multicolumn{2}{c}{\sq}
\\

\cmidrule(lr){3-6}
\cmidrule(lr){8-11}

{\twop}
&
& \bb{54.16} &
& \bb{0.08} & % 0.08259
&
& \bb{50.01} &
& \bb{0.73} & % 0.7268
\\

%\midrule

{\hdt}
&
& 68.79 & \small{(+21\%)}
& 19.67 & \small($238.2\times$)
&
& 92.41 & \small{(+46\%)}
& 17.26 & \small($23.7\times$)
\\

%{\rdfx}
%&
%& 87.42 & \small{(+38\%)}
%&  0.08 & \small($1.0\times$) % 0.08085
%&
%& 78.99 & \small{(+37\%)}
%&  0.67 & \small($0.9\times$) % 0.6655
%\\

{\triplebit}
&
& 129.49   & \small{(+58\%)}
&   0.34   & \small($4.1\times$) % 0.3419
&
& 111.93   & \small{(+55\%)}
&   4.01   & \small($5.5\times$) % 4.0137
\\

\bottomrule
\end{tabular}

}
\label{tab:sparql}
\end{table}

%As it is standard, the integer IDs assigned to each triple component are the lexicographic positions of the corresponding strings in the dictionaries.

%\return
%During our experiments we determined that the most effective index in terms of space of representation as well as the most efficient in terms of selection patterns resolution is the {\hdt} index, confirming the results shown by the corresponding authors~\cite{HDT1,HDT2}.
%Therefore, this is the index we compare against in the subsequent sections.
%We summarize the performance of this index in Table~\ref{tab:hdt} by using the C++ implementation made available by the authors at \url{https://github.com/rdfhdt/hdt-cpp}.
%The results hold for the solution called \emph{focused on querying} (FoQ)~\cite{HDT2}, that includes two additional indexes to speed up the execution of the different triple selection patterns.

%\giulio{{\bitmat} is the version 0.1 of the software (latest) and the indexes are built with the flag \texttt{COMPRESS\_FOLDED\_ARRAY} as recommended by the author Medha Atre (personal communication).}
%
%\giulio{The space for {\rdfx} does not include the space for the three additional permutations SOP, OPS and PSO, as well as the space of the bigram aggregate indexes.}

\section{Conclusions and Future Work}\label{sec:conclusions}
In this work we have proposed compressed indexes for the storage and search of large
RDF datasets, delivering a remarkably improved effectiveness/efficiency trade-off
against existing solutions.
In particular, the extensive experimentation provided has shown that
our best trade-off configuration reduces storage requirements by
30 -- 60\%
and provides 2 -- 81$\times$ better efficiency,
on real-world RDF datasets with up to 2 billions of triples.

Future work could target the two related problems mentioned in Section~\ref{sec:intro},
that is: (1) providing an efficient \emph{string dictionary} data structure and
(2) devising a novel \emph{query planning} algorithm.
%Another interesting research direction would be the one of
%supporting dynamic updates to the indexing data structure without
%compromising space effectiveness and retrieval performance.

\bibliographystyle{IEEEtran}
\bibliography{IEEEabrv,bibliography}

\begin{thebibliography}{10}
\providecommand{\url}[1]{#1}
\csname url@samestyle\endcsname
\providecommand{\newblock}{\relax}
\providecommand{\bibinfo}[2]{#2}
\providecommand{\BIBentrySTDinterwordspacing}{\spaceskip=0pt\relax}
\providecommand{\BIBentryALTinterwordstretchfactor}{4}
\providecommand{\BIBentryALTinterwordspacing}{\spaceskip=\fontdimen2\font plus
\BIBentryALTinterwordstretchfactor\fontdimen3\font minus
  \fontdimen4\font\relax}
\providecommand{\BIBforeignlanguage}[2]{{%
\expandafter\ifx\csname l@#1\endcsname\relax
\typeout{** WARNING: IEEEtran.bst: No hyphenation pattern has been}%
\typeout{** loaded for the language `#1'. Using the pattern for}%
\typeout{** the default language instead.}%
\else
\language=\csname l@#1\endcsname
\fi
#2}}
\providecommand{\BIBdecl}{\relax}
\BIBdecl

\bibitem{semWeb}
T.~Berners-Lee, J.~Hendler, and O.~Lassila, ``The semantic web,''
  \emph{Scientific American}, vol. 284, no.~5, pp. 34--43, 2001.

\bibitem{RDFSurvey}
M.~Wylot, M.~Hauswirth, P.~Cudr{\'e}-Mauroux, and S.~Sakr, ``Rdf data storage
  and query processing schemes: A survey,'' \emph{ACM Computing Surveys
  (CSUR)}, vol.~51, no.~4, p.~84, 2018.

\bibitem{RDFSurvey2}
M.~T. \"{O}zsu, ``A survey of rdf data management systems,'' \emph{Front.
  Comput. Sci.}, vol.~10, no.~3, pp. 418--432, Jun. 2016.

\bibitem{RDF3x}
T.~Neumann and G.~Weikum, ``The rdf-3x engine for scalable management of rdf
  data,'' \emph{The VLDB Journal—The International Journal on Very Large Data
  Bases}, vol.~19, no.~1, pp. 91--113, 2010.

\bibitem{RDF3xjournal}
------, ``x-rdf-3x: fast querying, high update rates, and consistency for rdf
  databases,'' \emph{Proceedings of the VLDB Endowment}, vol.~3, no. 1-2, pp.
  256--263, 2010.

\bibitem{TripleBit}
P.~Yuan, P.~Liu, B.~Wu, H.~Jin, W.~Zhang, and L.~Liu, ``Triplebit: a fast and
  compact system for large scale rdf data,'' \emph{Proceedings of the VLDB
  Endowment}, vol.~6, no.~7, pp. 517--528, 2013.

\bibitem{paulheim2013type}
H.~Paulheim and C.~Bizer, ``Type inference on noisy rdf data,'' in
  \emph{International semantic web conference}.\hskip 1em plus 0.5em minus
  0.4em\relax Springer, 2013, pp. 510--525.

\bibitem{subercaze2016inferray}
J.~Subercaze, C.~Gravier, J.~Chevalier, and F.~Laforest, ``Inferray: fast
  in-memory rdf inference,'' \emph{Proceedings of the VLDB Endowment}, vol.~9,
  no.~6, pp. 468--479, 2016.

\bibitem{Hexastore}
C.~Weiss, P.~Karras, and A.~Bernstein, ``Hexastore: sextuple indexing for
  semantic web data management,'' \emph{Proceedings of the VLDB Endowment},
  vol.~1, no.~1, pp. 1008--1019, 2008.

\bibitem{HDT2}
M.~A. Mart{\'\i}nez-Prieto, M.~A. Gallego, and J.~D. Fern{\'a}ndez, ``Exchange
  and consumption of huge rdf data,'' in \emph{Extended Semantic Web
  Conference}.\hskip 1em plus 0.5em minus 0.4em\relax Springer, 2012, pp.
  437--452.

\bibitem{BitMat}
M.~Atre, V.~Chaoji, M.~J. Zaki, and J.~A. Hendler, ``Matrix bit loaded: a
  scalable lightweight join query processor for rdf data,'' in
  \emph{Proceedings of the 19th international conference on World wide
  web}.\hskip 1em plus 0.5em minus 0.4em\relax ACM, 2010, pp. 41--50.

\bibitem{AbadiMMH09}
D.~J. Abadi, A.~Marcus, S.~Madden, and K.~Hollenbach, ``Sw-store: a vertically
  partitioned {DBMS} for semantic web data management,'' \emph{{VLDB} J.},
  vol.~18, no.~2, pp. 385--406, 2009.

\bibitem{sankar2014efficient}
S.~Sankar, M.~Singh, A.~Sayed, and J.~A. Bani-Younis, ``An efficient and
  scalable rdf indexing strategy based on b-hashed-bitmap algorithm using
  cuda,'' \emph{International Journal of Computer Applications}, vol. 104,
  no.~7, 2014.

\bibitem{brisaboa2009k}
N.~R. Brisaboa, S.~Ladra, and G.~Navarro, ``$k^2$-trees for compact web graph
  representation,'' in \emph{International Symposium on String Processing and
  Information Retrieval}.\hskip 1em plus 0.5em minus 0.4em\relax Springer,
  2009, pp. 18--30.

\bibitem{sadakane2003new}
K.~Sadakane, ``New text indexing functionalities of the compressed suffix
  arrays,'' \emph{Journal of Algorithms}, vol.~48, no.~2, pp. 294--313, 2003.

\bibitem{alvarez2015compressed}
S.~{\'A}lvarez-Garc{\'\i}a, N.~Brisaboa, J.~D. Fern{\'a}ndez, M.~A.
  Mart{\'\i}nez-Prieto, and G.~Navarro, ``Compressed vertical partitioning for
  efficient rdf management,'' \emph{Knowledge and Information Systems},
  vol.~44, no.~2, pp. 439--474, 2015.

\bibitem{brisaboa2015compact}
N.~R. Brisaboa, A.~Cerdeira-Pena, A.~Farina, and G.~Navarro, ``A compact rdf
  store using suffix arrays,'' in \emph{International Symposium on String
  Processing and Information Retrieval}.\hskip 1em plus 0.5em minus 0.4em\relax
  Springer, 2015, pp. 103--115.

\bibitem{cerdeira2016self}
A.~Cerdeira-Pena, A.~Farina, J.~D. Fern{\'a}ndez, and M.~A.
  Mart{\'\i}nez-Prieto, ``Self-indexing rdf archives,'' in \emph{2016 Data
  Compression Conference (DCC)}.\hskip 1em plus 0.5em minus 0.4em\relax IEEE,
  2016, pp. 526--535.

\bibitem{farina2012word}
A.~Fari{\~n}a, N.~R. Brisaboa, G.~Navarro, F.~Claude, {\'A}.~S. Places, and
  E.~Rodr{\'\i}guez, ``Word-based self-indexes for natural language text,''
  \emph{ACM Transactions on Information Systems (TOIS)}, vol.~30, no.~1, p.~1,
  2012.

\bibitem{HDT1}
J.~D. Fern{\'a}ndez, M.~A. Mart{\'\i}nez-Prieto, and C.~Gutierrez, ``Compact
  representation of large rdf data sets for publishing and exchange,'' in
  \emph{International Semantic Web Conference}.\hskip 1em plus 0.5em minus
  0.4em\relax Springer, 2010, pp. 193--208.

\bibitem{grossi2003}
R.~Grossi, A.~Gupta, and J.~S. Vitter, ``High-order entropy-compressed text
  indexes,'' in \emph{Proceedings of the fourteenth annual ACM-SIAM symposium
  on Discrete algorithms}, 2003, pp. 841--850.

\bibitem{CureBRF14}
O.~Cur{\'{e}}, G.~Blin, D.~Revuz, and D.~C. Faye, ``Waterfowl: {A} compact,
  self-indexed and inference-enabled immutable {RDF} store,'' in
  \emph{Proceedings of the 11th International Conference on The Semantic Web:
  Trends and Challenges ({ESWC})}, 2014, pp. 302--316.

\bibitem{auer2007dbpedia}
S.~Auer, C.~Bizer, G.~Kobilarov, J.~Lehmann, R.~Cyganiak, and Z.~Ives,
  ``Dbpedia: A nucleus for a web of open data,'' pp. 722--735, 2007.

\bibitem{survey}
\BIBentryALTinterwordspacing
G.~E. Pibiri and R.~Venturini, ``Techniques for inverted index compression,''
  \emph{CoRR}, vol. abs/1908.10598, 2019. [Online]. Available:
  \url{http://arxiv.org/abs/1908.10598}
\BIBentrySTDinterwordspacing

\bibitem{zobel2006inverted}
J.~Zobel and A.~Moffat, ``Inverted files for text search engines,'' \emph{ACM
  computing surveys (CSUR)}, vol.~38, no.~2, p.~6, 2006.

\bibitem{pibiri2018inverted}
G.~E. Pibiri and R.~Venturini, ``Inverted index compression,''
  \emph{Encyclopedia of Big Data Technologies}, pp. 1--8, 2018.

\bibitem{elias}
P.~Elias, ``Efficient storage and retrieval by content and address of static
  files,'' \emph{Journal of the ACM (JACM)}, vol.~21, no.~2, pp. 246--260,
  1974.

\bibitem{fano}
R.~M. Fano, ``On the number of bits required to implement an associative
  memory,'' \emph{Memorandum 61, Computer Structures Group, MIT}, 1971.

\bibitem{ottaviano2014partitioned}
G.~Ottaviano and R.~Venturini, ``Partitioned elias-fano indexes,'' in
  \emph{Proceedings of the 37th international ACM SIGIR conference on Research
  \& development in information retrieval}.\hskip 1em plus 0.5em minus
  0.4em\relax ACM, 2014, pp. 273--282.

\bibitem{pibiri2017efficient}
G.~E. Pibiri and R.~Venturini, ``Efficient data structures for massive n-gram
  datasets,'' in \emph{Proceedings of the 40th International ACM SIGIR
  Conference on Research and Development in Information Retrieval}.\hskip 1em
  plus 0.5em minus 0.4em\relax ACM, 2017, pp. 615--624.

\bibitem{Pibiri2019}
------, ``Handling massive n-gram datasets efficiently,'' \emph{ACM Trans. Inf.
  Syst.}, vol.~37, no.~2, pp. 25:1--25:41, Feb. 2019.

\bibitem{thiel1972program}
L.~H. Thiel and H.~Heaps, ``Program design for retrospective searches on large
  data bases,'' \emph{Information Storage and Retrieval (ISR)}, vol.~8, no.~1,
  pp. 1--20, 1972.

\bibitem{maskedvb}
J.~Plaisance, N.~Kurz, and D.~Lemire, ``Vectorized {VB}yte decoding,'' in
  \emph{International Symposium on Web Algorithms (iSWAG)}, 2015.

\bibitem{dblp}
DBLP, ``Computer science bibliography,'' \url{http://www.rdfhdt.org/datasets},
  2017.

\bibitem{geo}
geonames.org, ``Geonames,'' \url{http://www.rdfhdt.org/datasets}, 2012.

\bibitem{watdiv}
G.~Alu{\c{c}}, O.~Hartig, M.~T. {\"O}zsu, and K.~Daudjee, ``Diversified stress
  testing of rdf data management systems,'' in \emph{International Semantic Web
  Conference}.\hskip 1em plus 0.5em minus 0.4em\relax Springer, 2014, pp.
  197--212.

\bibitem{lubm}
LUBM, ``Lehigh university benchmark,''
  \url{http://swat.cse.lehigh.edu/projects/lubm}, 2019.

\bibitem{freebase}
Google, ``Freebase data dumps,'' \url{https://developers.google.com/freebase},
  2013.

\end{thebibliography}

\vspace{-1.5cm}
\begin{IEEEbiography}[{\includegraphics[width=1in,height=1.25in,clip,keepaspectratio]{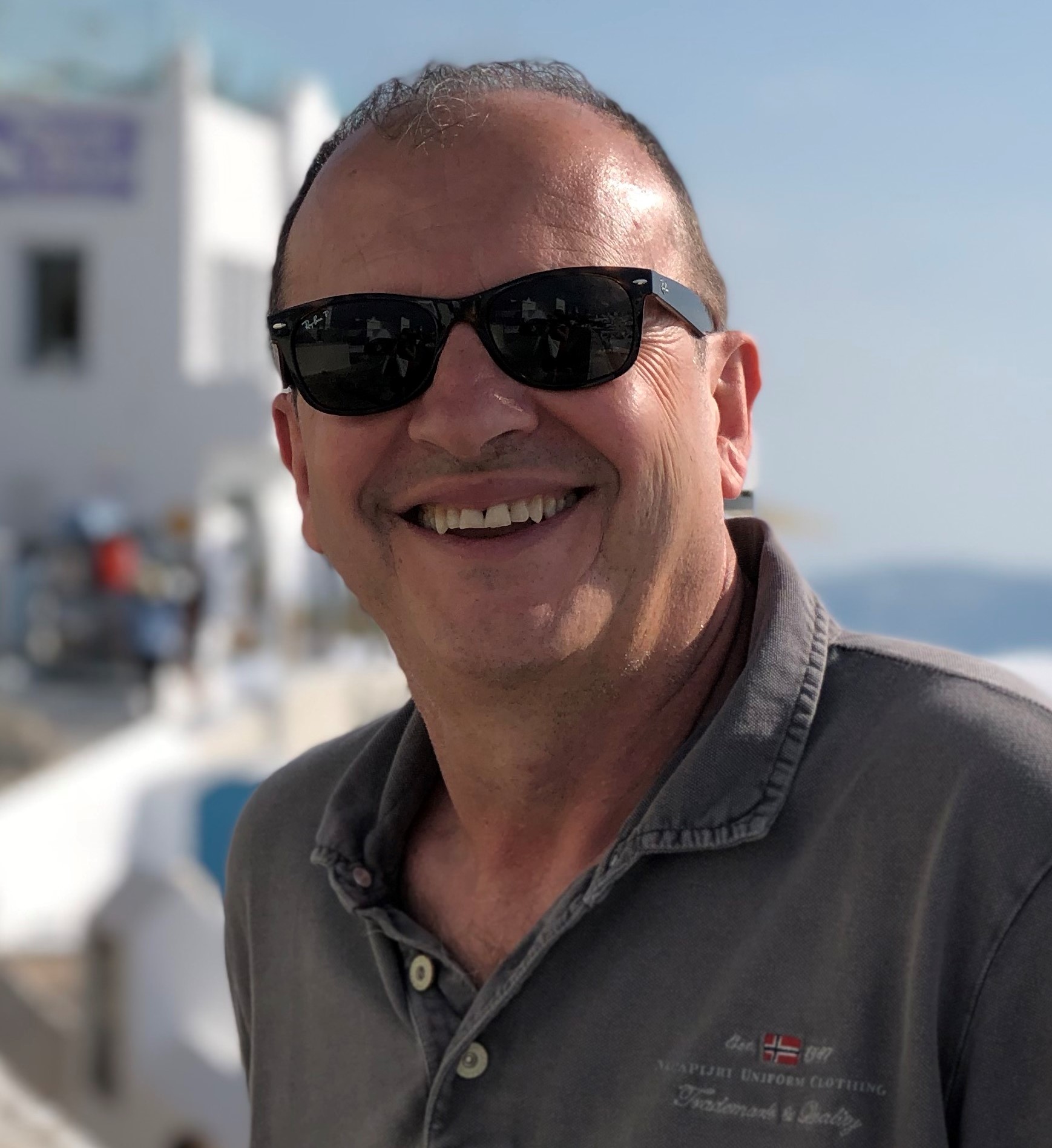}}]{Raffaele Perego}(\url{http://hpc.isti.cnr.it/~raffaele/})
is a senior researcher at ISTI-CNR, where he leads the High Performance Computing Lab (http://hpc.isti.cnr.it). His main research interests include large-scale information systems, information retrieval, data mining, and machine learning. He co-authored more than 150 papers on these topics published in journals and proceedings of international conferences.
\end{IEEEbiography}

\vspace{-1.5cm}
\begin{IEEEbiography}[{\includegraphics[width=1in,height=1.25in,clip,keepaspectratio]{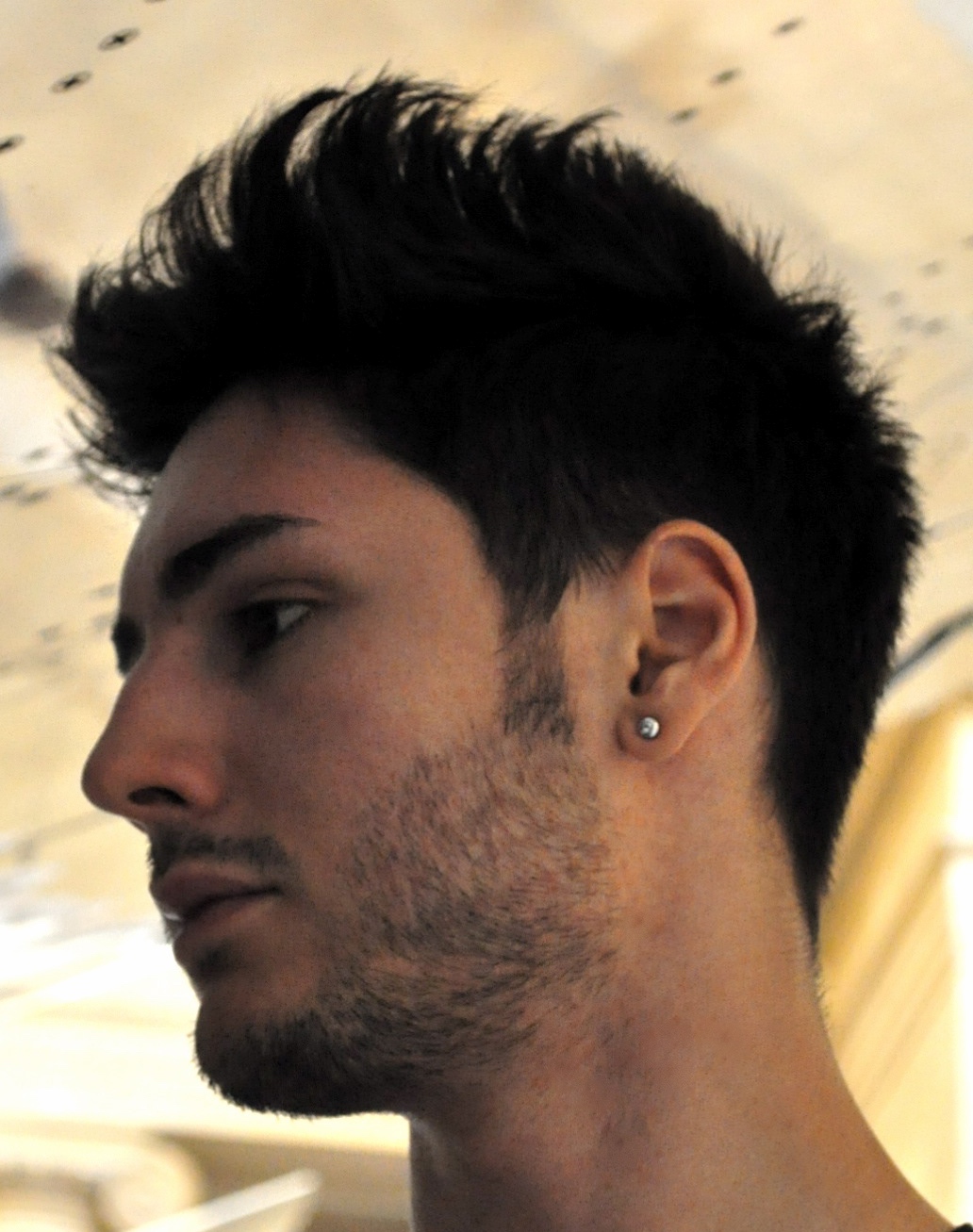}}]{Giulio Ermanno Pibiri}(\url{http://pages.di.unipi.it/pibiri}) is a postdoctoral researcher at ISTI-CNR, High Performance Computing Lab.
He received a Ph.D. in Computer Science from the University of Pisa in 2019.
His research interests involve data compression algorithms for indexing massive datasets, data structures and information retrieval with focus on efficiency.
\end{IEEEbiography}

\vspace{-1.3cm}
\begin{IEEEbiography}
[{\includegraphics[width=1in,height=1.25in,clip,keepaspectratio]{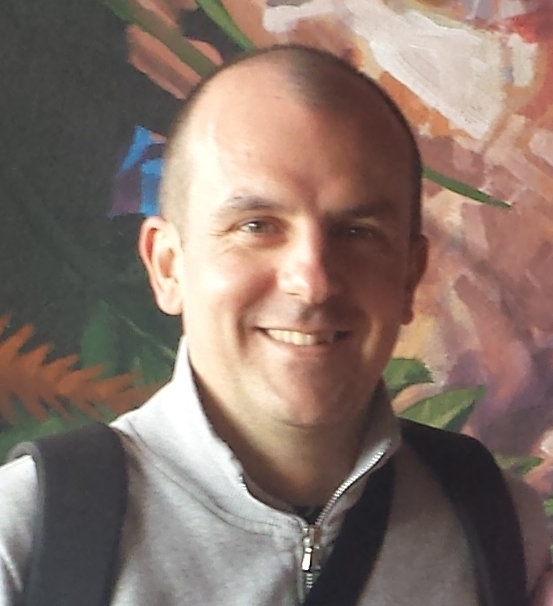}}] {Rossano Venturini}(\url{http://pages.di.unipi.it/rossano}) is Associate Professor of Computer Science at the University of Pisa. He received his Ph.D. from the University of Pisa in 2010. His research interests are mainly focused on the design and the analysis of algorithms and data structures with focus on indexing and searching large textual collections.
He won two Best Paper Awards at ACM SIGIR in 2014 and 2015.
\end{IEEEbiography} 

\end{document}